\documentclass[aps,preprint]{revtex4}%
\usepackage{amsfonts}
\usepackage{amsmath}
\usepackage{amssymb}
\usepackage{subfigure}
\usepackage{graphicx}%
\begin{document}
\preprint{CTP-SCU/2017005}
\title{Testing Holographic Conjectures of Complexity with Born-Infeld Black Holes }
\author{Jun Tao}
\email{taojun@scu.edu.cn}
\author{Peng Wang}
\email{pengw@scu.edu.cn}
\author{Haitang Yang}
\email{hyanga@scu.edu.cn}
\affiliation{Center for Theoretical Physics, College of Physical Science and Technology,
Sichuan University, Chengdu, 610064, China}

\begin{abstract}
In this paper, we use Born-Infeld black holes to test two recent holographic
conjectures of complexity, the \textquotedblleft Complexity =
Action\textquotedblright\ (CA)\ duality and \textquotedblleft Complexity =
Volume 2.0\textquotedblright\ (CV)\ duality. The complexity of a boundary
state is identified with the action of the Wheeler-deWitt patch in CA duality,
while this complexity is identified with the spacetime volume of the WdW patch
in CV duality. In particular, we check whether the Born-Infeld black holes
violate the Lloyd bound: $\mathcal{\dot{C}\leq}\frac{2}{\pi\hbar}\left[
\left(  M-Q\Phi\right)  -\left(  M-Q\Phi\right)  _{\text{gs}}\right]  $, where
gs stands for the ground state for a given electrostatic potential. We find
that the ground states are either some extremal black hole or regular
spacetime with nonvanishing charges. Near extremality, the Lloyd bound is
violated in both dualities. Near the charged regular spacetime, this bound is
satisfied in CV\ duality but violated in CA\ duality. When moving away from
the ground state on a constant potential curve, the Lloyd bound tend to be
saturated from below in CA duality, while $\mathcal{\dot{C}}$ is $\pi/2$ times
as large as the Lloyd bound in CV\ duality.

\end{abstract}
\maketitle
\tableofcontents



\section{Introduction}

Through gauge/gravity duality, concepts from quantum information theory have
driven major advances in our understanding of quantum field theory and quantum
gravity. For example, the holographic entanglement entropy
\cite{IN-Ryu:2006bv,IN-Ryu:2006ef} has been currently receiving considerable
attentions in the ongoing research. Recently inspired by the observation that
the size of the Einstein-Rosen bridge (ERB) grows linearly at late times, it
was conjectured
\cite{IN-Susskind:2014rva,IN-Stanford:2014jda,IN-Susskind:2014jwa,IN-Susskind:2014moa}
that quantum complexity of a boundary state is dual to the volume of the
maximal spatial slice crossing the ERB anchored at the boundary state. Roughly
speaking, the complexity $\mathcal{C}$ of a state is the minimum number of
quantum gates to prepare this state from a reference state
\cite{IN-Wat:2009,IN-Huang:2015,IN-Osb:2011}. However, one of unappealing
features of this proposal is that there is an ambiguity in choosing a length
scale in the bulk geometry, which provides some motivations to introduce the
\textquotedblleft Complexity = Action\textquotedblright\ (CA)\ duality
\cite{IN-Brown:2015bva,IN-Brown:2015lvg}.

In CA duality, the complexity of a boundary state is identified with the
action of the Wheeler-DeWitt (WdW) patch in the bulk:%
\begin{equation}
\mathcal{C=}\frac{S_{\text{WdW}}}{\pi\hbar},
\end{equation}
where the WdW patch can be defined as the domain of dependence of any Cauchy
surface anchored at the boundary state. After the original calculations of
$S_{\text{WdW}}$ in \cite{IN-Brown:2015lvg}, a detailed analysis was carried
out in \cite{IN-Lehner:2016vdi}, of the contributions to the action of some
subregion from a null segment and a joint at which a null segment is joined to
another segment. It is interesting to note that although the two approaches
used in \cite{IN-Brown:2015lvg} and \cite{IN-Lehner:2016vdi} are different,
the results for $dS_{\text{WdW}}/dt$ at late times of the AdS Schwarzschild
and Reissner-Nordstrom (RN) AdS black holes turn out to be the same. A
possible explanation was given in \cite{IN-Lehner:2016vdi}.

Similar to the holographic entanglement entropy, the holographic complexity in
CA duality is divergent, which is related to the infinite volume near the
boundary of AdS space. The divergent terms were considered in
\cite{IN-Carmi:2016wjl,IN-Reynolds:2016rvl,IN-Kim:2017lrw}, which showed that
these terms could be written as local integrals of boundary geometry. This
implies that the divergence comes from the UV degrees of freedom in the field
theory. On the other hand, there are two finite quantities associated with the
complexity, which can be calculated without first obtaining these divergent
terms. The first one is the \textquotedblleft complexity of
formation\textquotedblright\ \cite{IN-Chapman:2016hwi}, which is the
difference of the complexity between a particular black hole and a vacuum AdS
spacetime. The second one is the rate of complexity at late times,
$\mathcal{\dot{C}}$. If CA duality is correct, $\mathcal{\dot{C}}$ should
saturate the Lloyd bound \cite{IN-Llo:2000}. The Lloyd bound is the
conjectured complexity growth bound, which states that $\mathcal{\dot{C}}$
should be bounded by the energy \cite{IN-Brown:2015lvg}:%
\begin{equation}
\mathcal{\dot{C}\leq}\frac{2E}{\pi\hbar}.
\end{equation}
For a black hole, $E$ is its mass $M$, and the Lloyd bound then reads%
\begin{equation}
\mathcal{\dot{C}\leq}\frac{2M}{\pi\hbar}.\label{eq:LBound}%
\end{equation}
As noted in \cite{IN-Brown:2015lvg}, the rate of the complexity of a neutral
black hole is faster than that of a charged black hole since the existence of
conserved charges could put constraints on the system. That implies that the
Lloyd bound can be generalized for a charged black hole with the charge $Q$
and potential at the horizon $\Phi$:%
\begin{equation}
\mathcal{\dot{C}\leq}\frac{2}{\pi\hbar}\left[  \left(  M-Q\Phi\right)
-\left(  M-Q\Phi\right)  _{\text{gs}}\right]  ,\label{eq:LBoundQ}%
\end{equation}
where $\left(  M-Q\Phi\right)  _{\text{gs}}$ is $M-Q\Phi$ calculated in the
ground state. A similar bound can also be given for rotating black holes
\cite{IN-Brown:2015lvg}.

The rate of complexity in CA duality has been considered in several examples.
In \cite{IN-Brown:2015lvg}, it showed that neutral black holes, rotating BTZ
black holes, and small RN AdS black holes saturated the corresponding Lloyd
bounds, while intermediate and large RN AdS black holes violated the bound
$\left(  \ref{eq:LBoundQ}\right)  $. Later, it was pointed out
\cite{IN-Cai:2016xho,IN-Couch:2016exn} that even the small RN AdS black holes
also violated the bound $\left(  \ref{eq:LBoundQ}\right)  $. The WdW patch
action growth of RN AdS black holes, (charged) rotating BTZ black holes, AdS
Kerr black holes, and (charged) Gauss-Bonnet black holes were calculated in
\cite{IN-Cai:2016xho}. The action growth was also discussed in case of massive
gravities \cite{IN-Pan:2016ecg} and higher derivative gravities
\cite{IN-Alishahiha:2017hwg}. A general case was considered in
\cite{IN-Huang:2016fks}, and it was proved that the action growth rate equals
the difference of the generalized enthalpy at the outer and inner horizons.
While this paper is in preparation, a preprint \cite{IN-Cai:2017sjv} appeared
calculating the action growth of Born-Infeld black holes, charged dilaton
black holes, and charged black holes with phantom Maxwell field in AdS space.
It also showed there that a Born-Infeld AdS black hole with a single horizon
and a charged dilaton AdS black hole satisfied the Lloyd bound $\left(
\ref{eq:LBound}\right)  $, while for the charged black hole with a phantom
Maxwell field, this bound was violated.

Noting that the thermodynamic volume was related to the linear growth of the
WdW patch at late times, Couch \textit{et. al.} proposed \textquotedblleft
Complexity = Volume 2.0\textquotedblright\ duality in \cite{IN-Couch:2016exn}.
In \textquotedblleft Complexity = Volume 2.0\textquotedblright\ (CV)\ duality,
the complexity is identified with the spacetime volume of the WdW patch. It
was found that the Lloyd bound $\left(  \ref{eq:LBoundQ}\right)  $ was
violated in both CA and CV dualities for RN AdS black holes near extremality.
However if the ground state was an empty AdS space, this bound was violated in
CA duality but satisfied in CV duality. In what follows, let $\mathcal{C}%
_{A}/\mathcal{C}_{V}$ denote the complexity calculated in CA/CV duality.

In this paper, we will check whether the generalized Lloyd bound $\left(
\ref{eq:LBoundQ}\right)  $ is violated for the Born-Infeld AdS black holes in
CA and CV dualities. The remainder of our paper is organized as follows. In
section \ref{Sec:BIABH}, we discuss some properties of Born-Infeld AdS black
holes, which could have a naked singularity, a single horizon, or two horizons
depending on their parameters. The phase diagrams for these black holes are
obtained. In section \ref{Sec:HCC}, we consider the Lloyd bound for the
Born-Infeld AdS black holes in CA/CV dualities. In section \ref{Sec:Con}, we
conclude with a brief discussion of our results. In the appendix, we employ
the approach in \cite{IN-Lehner:2016vdi} to calculate action growth for
$\left(  d+1\right)  $-dimensional Born-Infeld AdS black holes with
hyperbolic, planar, and spherical horizons.

\section{Born-Infeld AdS Black Holes}

\label{Sec:BIABH}

In this section, we will consider the black hole solutions of
Einstein-Born-Infeld action in $\left(  d+1\right)  $ dimension $\left(
d\geq3\right)  $ with a negative cosmological constant $\Lambda=-\frac
{d\left(  d-1\right)  }{L^{2}}$. The action of Einstein gravity and
Born-Infeld field reads%
\begin{equation}
S=\int_{\mathcal{M}}d^{d+1}x\sqrt{-g}\left(  R+\frac{d\left(  d-1\right)
}{L^{2}}\right)  +\int_{\mathcal{M}}d^{d+1}x\sqrt{-g}L\left(  F\right)  ,
\end{equation}
where we take $16\pi G=1$ for simplicity, $L\left(  F\right)  $ is given by%
\begin{equation}
L\left(  F\right)  =4\beta^{2}\left(  1-\sqrt{1+\frac{F^{\mu\nu}F_{\mu\nu}%
}{2\beta^{2}}}\right)  ,
\end{equation}
and, $\beta$ is the Born-Infeld parameter. When $\beta\rightarrow\infty$, the
Lagrangian of Born-Infeld field $L\left(  F\right)  $ becomes that of standard
Maxwell field, $L\left(  F\right)  =-F^{\mu\nu}F_{\mu\nu}.$ The static black
hole solution was obtained in \cite{BIABH-Dey:2004yt,BIABH-Cai:2004eh}:%
\begin{align}
ds^{2} &  =-f\left(  r\right)  dt^{2}+\frac{dr^{2}}{f\left(  r\right)  }%
+r^{2}d\Sigma_{k,d-1}^{2},\nonumber\\
F^{rt} &  =\frac{\sqrt{\left(  d-1\right)  \left(  d-2\right)  }\beta q}%
{\sqrt{2\beta^{2}r^{2d-2}+\left(  d-1\right)  \left(  d-2\right)  q^{2}}%
},\label{eq:BIBH}%
\end{align}
where%
\begin{align}
f\left(  r\right)   &  =k-\frac{m}{r^{d-2}}+\left[  \frac{4\beta^{2}}{d\left(
d-1\right)  }+\frac{1}{L^{2}}\right]  r^{2}-\frac{2\sqrt{2}\beta}{d\left(
d-1\right)  r^{d-3}}\sqrt{2\beta^{2}r^{2d-2}+\left(  d-1\right)  \left(
d-2\right)  q^{2}}\nonumber\\
&  +\frac{2\left(  d-1\right)  q^{2}}{dr^{2d-4}}\text{ }_{2}F_{1}\left[
\frac{d-2}{2d-2},\frac{1}{2},\frac{3d-4}{2d-2},-\frac{\left(  d-1\right)
\left(  d-2\right)  q^{2}}{2\beta^{2}r^{2d-2}}\right]  ,
\end{align}
and, $d\Sigma_{k,d-1}^{2}$ is the line element of the $\left(  d-1\right)
$-dimensional hypersurface with constant scalar curvature $\left(  d-1\right)
\left(  d-2\right)  k$ with $k=\left\{  -1,0,1\right\}  $. Note that the black
holes with $k=\left\{  -1,0,1\right\}  $ have hyperbolic, planar, and
spherical horizons. The mass $M$ and charge $Q$ of the Born-Infeld black hole
are given by, respectively,%
\begin{align}
M &  =\left(  d-1\right)  \Omega_{k,d-1}m,\nonumber\\
Q &  =\frac{\sqrt{\left(  d-1\right)  \left(  d-2\right)  }\Omega_{k,d-1}%
}{4\pi\sqrt{2}}q,
\end{align}
where $\Omega_{k,d-1}$ denotes the dimensionless volume of $d\Sigma
_{k,d-1}^{2}$. For $k=0$ and $-1$, one needs to introduce an infrared
regulator to produce a finite value of $\Omega_{k,d-1}$.

\begin{figure}[tb]
\begin{center}
\subfigure[{~ $k=1$ and $d=3$}]{
\includegraphics[width=0.48\textwidth]{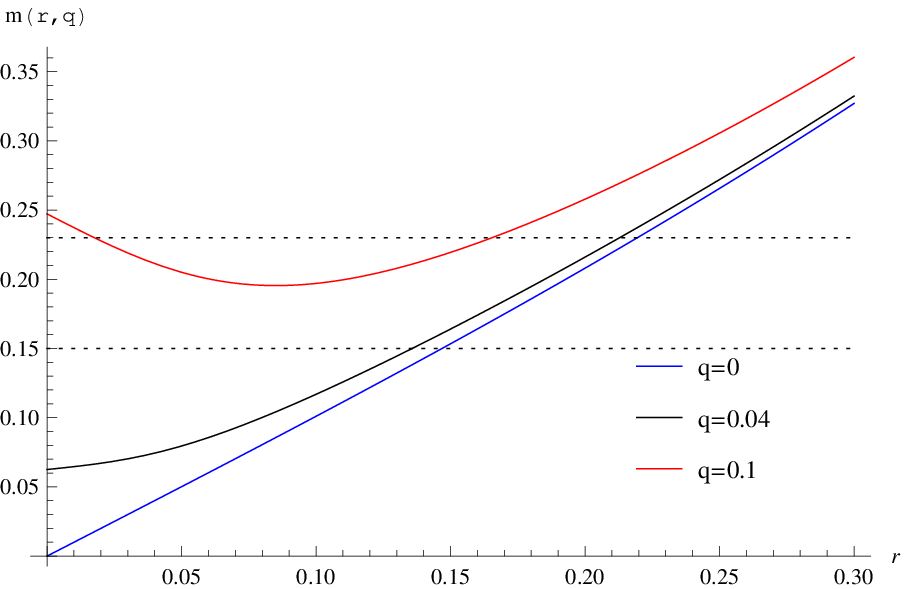}\label{fig:mr:a}}
\subfigure[{~$k=-1$ and $d=3$}]{
\includegraphics[width=0.48\textwidth]{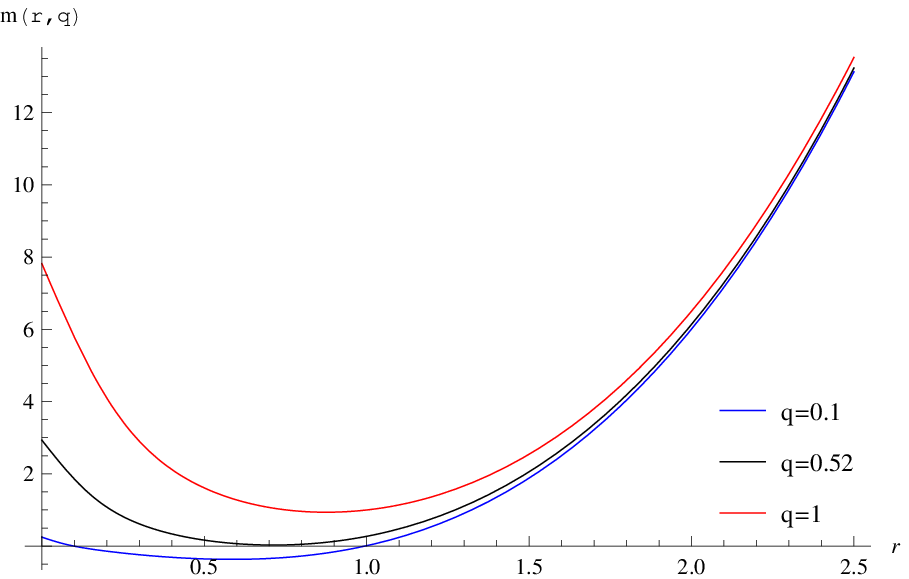}\label{fig:mr:b}}
\subfigure[{~ $k=0$ and $d=3$}]{
\includegraphics[width=0.48\textwidth]{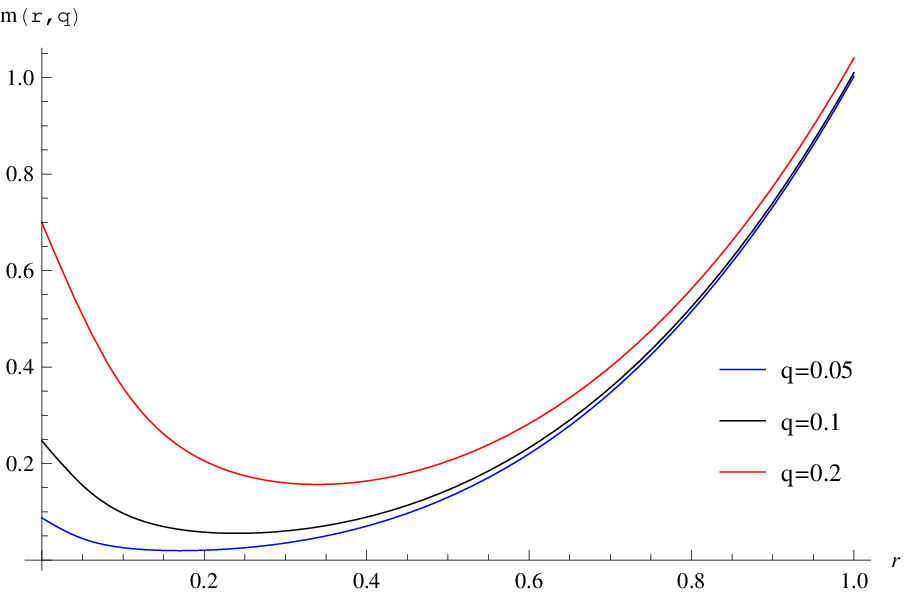}\label{fig:mr:c}}
\subfigure[{~ $k=1$ and $d=4$}]{
\includegraphics[width=0.48\textwidth]{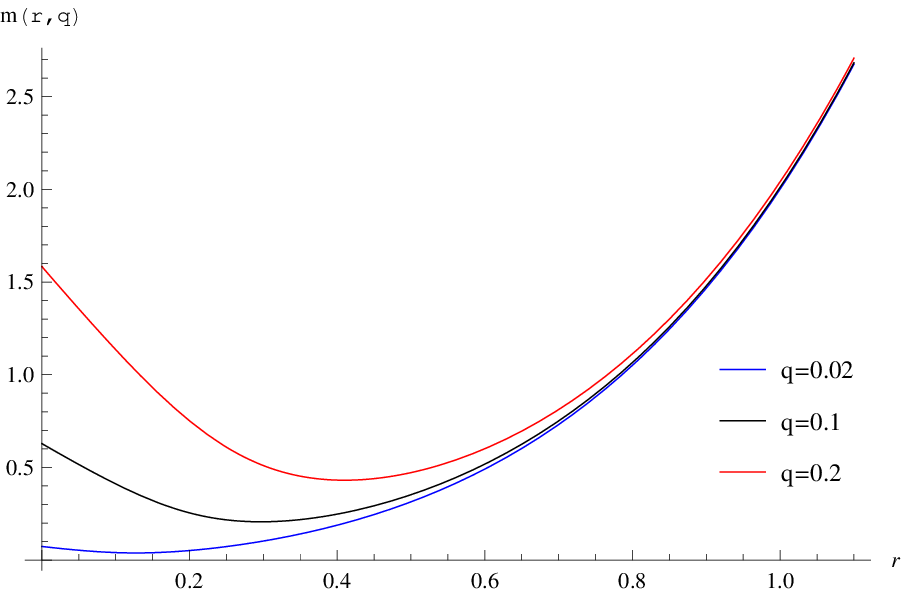}\label{fig:mr:d}}
\end{center}
\caption{Plots of $m\left(  r,q\right)  $ versus $r$ for different values of
$q$, where $L=1$ and $\beta=10$.}%
\label{fig:mr}%
\end{figure}

For the sake of calculating the action growth and thermodynamic volume of the
Born-Infeld black holes, we need to determine the number of their horizons.
Depending on the values of the parameters $q$ and $m$, the black holes could
possess a naked singularity at $r=0$, one, or two horizons. In fact, we could
define a $q$-dependent function
\begin{equation}
m\left(  r,q\right)  =r^{d-2}f\left(  r\right)  +m,
\end{equation}
which does not depend on the parameter $m$. For a given value of $m$, one
could solve $m\left(  r,q\right)  =m$ for the position of the horizon. The
derivative of $m\left(  r,q\right)  $ with respect to $r$ is%
\begin{equation}
\frac{dm\left(  r,q\right)  }{dr}=\left(  d-2\right)  r^{d-3}\left[
k+\frac{dr^{2}}{\left(  d-2\right)  L^{2}}-\frac{2q^{2}}{r^{d+1}\left(
r^{d-1}+\sqrt{r^{2d-2}+\frac{\left(  d-1\right)  \left(  d-2\right)  q^{2}%
}{2\beta^{2}}}\right)  }\right]  ,
\end{equation}
which is a strictly increasing function. When $r\rightarrow\infty$, $dm\left(
r,q\right)  /dr$ goes to $\infty$. In the limit $r\rightarrow0$, we find that%
\begin{align}
\frac{dm\left(  r,q\right)  }{dr}|_{r=0} &  =-\frac{2\sqrt{2}\beta q}{\left(
d-1\right)  }\sqrt{\left(  d-1\right)  \left(  d-2\right)  },\text{ for
}d>3\text{,}\nonumber\\
\frac{dm\left(  r,q\right)  }{dr}|_{r=0} &  =k-2\beta q\text{, for }d=3,
\end{align}
which shows that $dm\left(  r,q\right)  /dr|_{r=0}\geq0$ in the $k=1$, $d=3$,
and $\beta q\leq1/2$ case, and $dm\left(  r,q\right)  /dr|_{r=0}<0$ in the
other cases. When $\frac{dm\left(  r,q\right)  }{dr}|_{r=0}<0$, the equation
$dm\left(  r,q\right)  /dr=0$ has one and only solution $r_{e}\left(
q\right)  >0$, such that $dm\left(  r,q\right)  /dr|_{r=r_{e}\left(  q\right)
}=0$. Thus, there is an extremal black hole solution with the parameter
$m=m\left(  r_{e}\left(  q\right)  ,q\right)  $ and the horizon being at
$r=r_{e}\left(  q\right)  $. At $r=r_{e}\left(  q\right)  $, we obtain%
\begin{equation}
m\left(  r_{e}\left(  q\right)  ,q\right)  =\frac{2}{d}kr_{e}^{d-2}%
+\frac{2\left(  d-1\right)  q^{2}}{dr_{e}^{d-2}}\text{ }_{2}F_{1}\left[
\frac{d-2}{2d-2},\frac{1}{2},\frac{3d-4}{2d-2},-\frac{\left(  d-1\right)
\left(  d-2\right)  q^{2}}{2\beta^{2}r_{e}^{2d-2}}\right]  .
\end{equation}
When $k=0$ and $1$, $m\left(  r_{e}\left(  q\right)  ,q\right)  $ is always
positive. However for $k=-1$, $m\left(  r_{e}\left(  q\right)  ,q\right)  $
could be negative for some values of $q$. It is noteworthy that $m\left(
r_{e}\left(  q\right)  ,q\right)  $ exists for $q\geq\frac{1}{2\beta}$ in the
$k=1$, $d=3$, and $\beta q\leq1/2$ case, while $m\left(  r_{e}\left(
q\right)  ,q\right)  \,$\ exists for all values of $q$ in other cases.
Moreover, one finds that
\begin{equation}
m\left(  0,q\right)  =A\left(  q\right)  >0,
\end{equation}
where
\begin{equation}
A\left(  q\right)  \equiv\frac{2\left(  d-1\right)  q^{2}}{d}\frac
{\Gamma\left(  \frac{3d-4}{2d-2}\right)  \Gamma\left(  \frac{1}{2\left(
d-1\right)  }\right)  }{\sqrt{\pi}}\left[  \frac{2\beta^{2}}{\left(
d-1\right)  \left(  d-2\right)  q^{2}}\right]  ^{\frac{d-2}{2\left(
d-1\right)  }}.\label{eq:Aq}%
\end{equation}
Since $dm\left(  r,q\right)  /dr<0$ for $0<r<r_{e}$, one obtains $m\left(
r_{e}\left(  q\right)  ,q\right)  <m\left(  0,q\right)  =A\left(  q\right)  $.
In FIG. $\ref{fig:mr}$, we plot the function $m\left(  r,q\right)  $ against
$r$ for different values of $q$, where we take $L=1$ and $\beta=10$.

With the above results, we can discuss when the Born-Infeld black hole
solution $\left(  \ref{eq:BIBH}\right)  $ possesses a naked singularity, a
single horizon, or two horizons:

\begin{itemize}
\item Single Horizon: $m\geq A\left(  q\right)  $. For example, $\left\{
q=0.04\text{, }m=0.15\right\}  $ in FIG. $\ref{fig:mr:a}$

\item Two Horizons: $m\left(  r_{e}\left(  q\right)  ,q\right)  \leq
m<A\left(  q\right)  $. For example, $\left\{  q=0.1\text{, }m=0.23\right\}  $
in FIG. $\ref{fig:mr:a}$

\item Naked Singularity: $m<A\left(  q\right)  $ when $k=1$, $d=3$, and $\beta
q\leq1/2$; $m<m\left(  r_{e}\left(  q\right)  ,q\right)  $ in the other cases.
For example, $\left\{  q=0.1\text{, }m=0.15\right\}  $ in FIG. $\ref{fig:mr:a}%
$.
\end{itemize}

\begin{figure}[tb]
\begin{center}
\subfigure[{~ $k=1$ and $d=3$}]{
\includegraphics[width=0.48\textwidth]{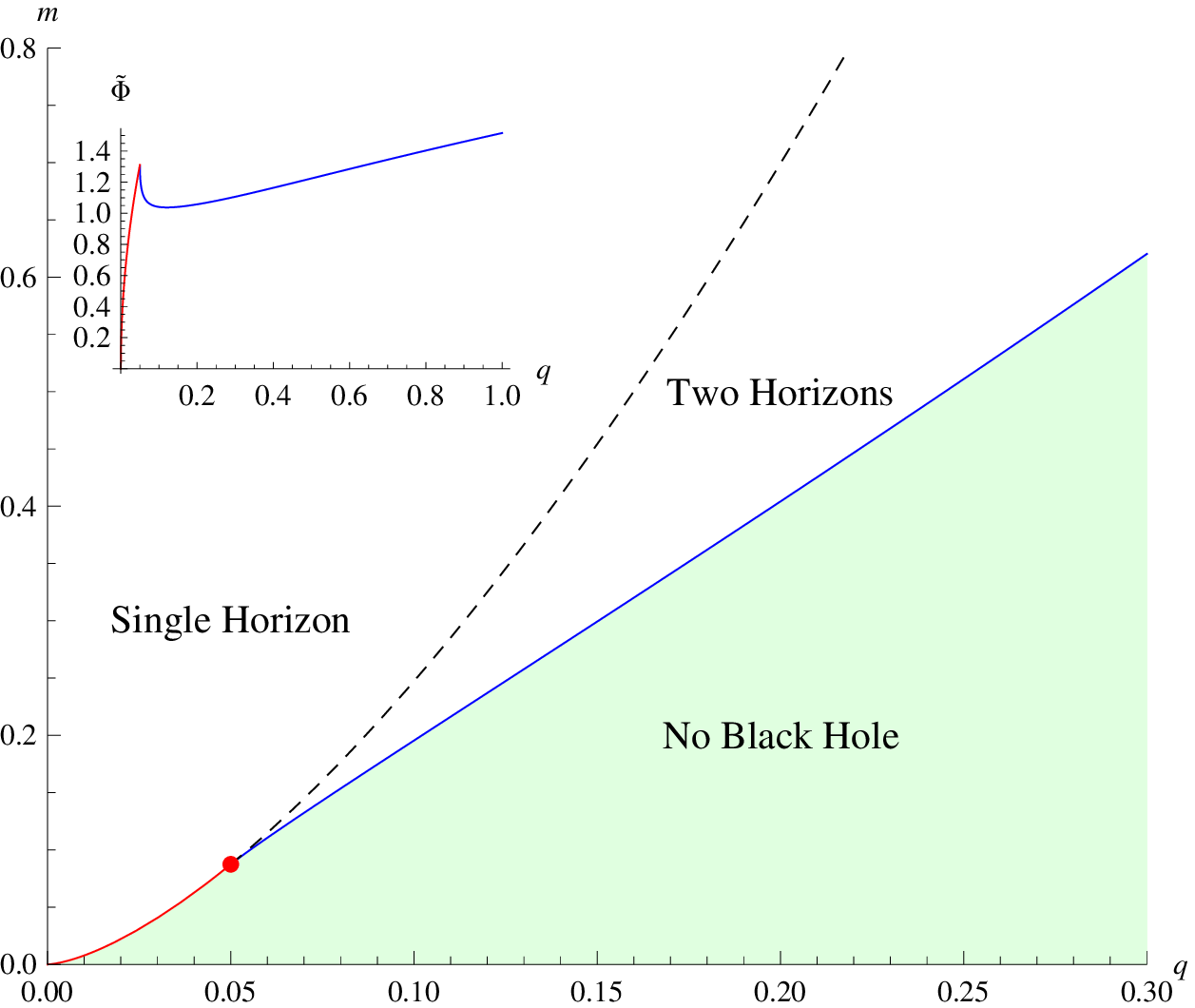}\label{fig:mq:a}}
\subfigure[{~$k=-1$ and $d=3$}]{
\includegraphics[width=0.48\textwidth]{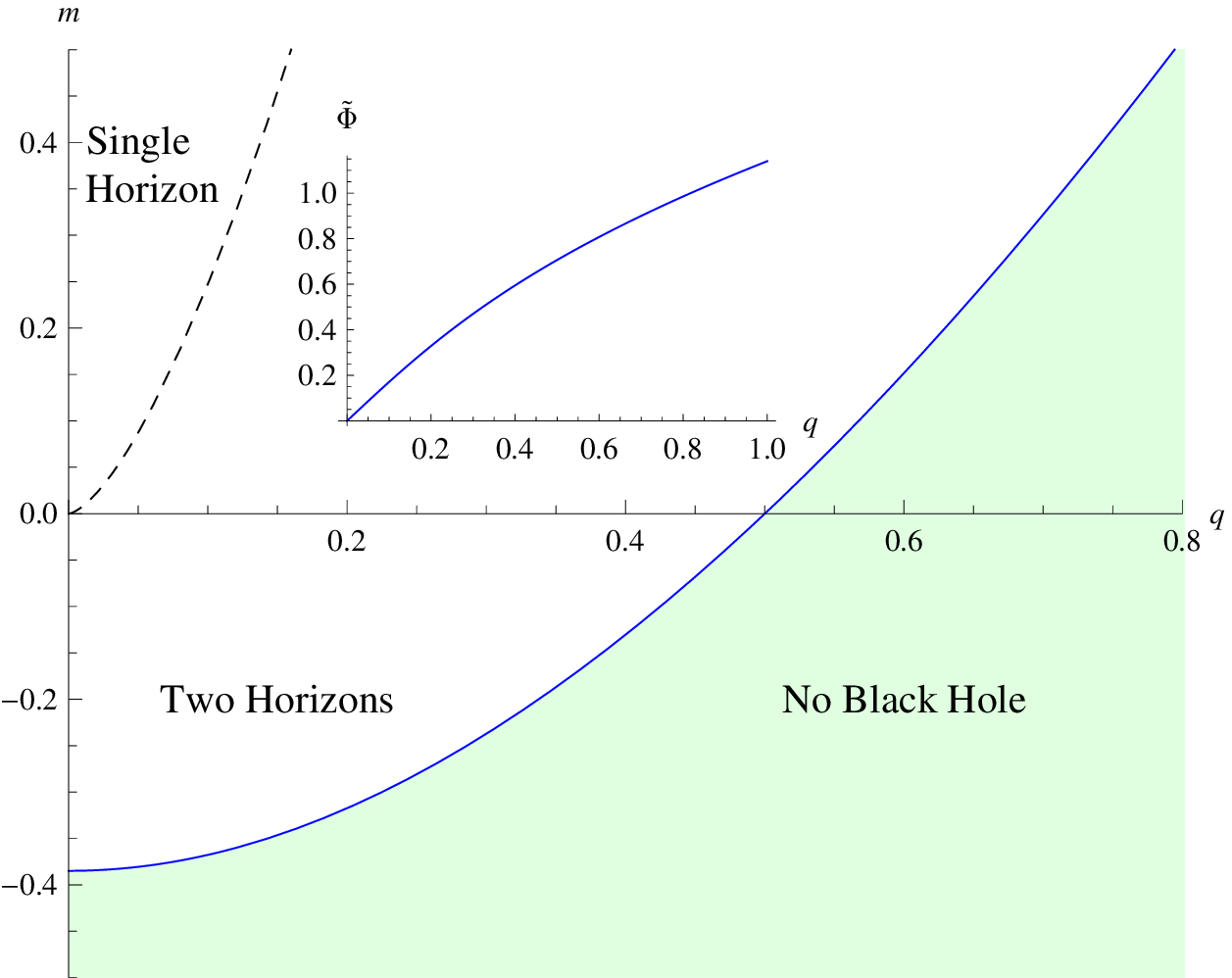}\label{fig:mq:b}}
\subfigure[{~ $k=0$ and $d=3$}]{
\includegraphics[width=0.48\textwidth]{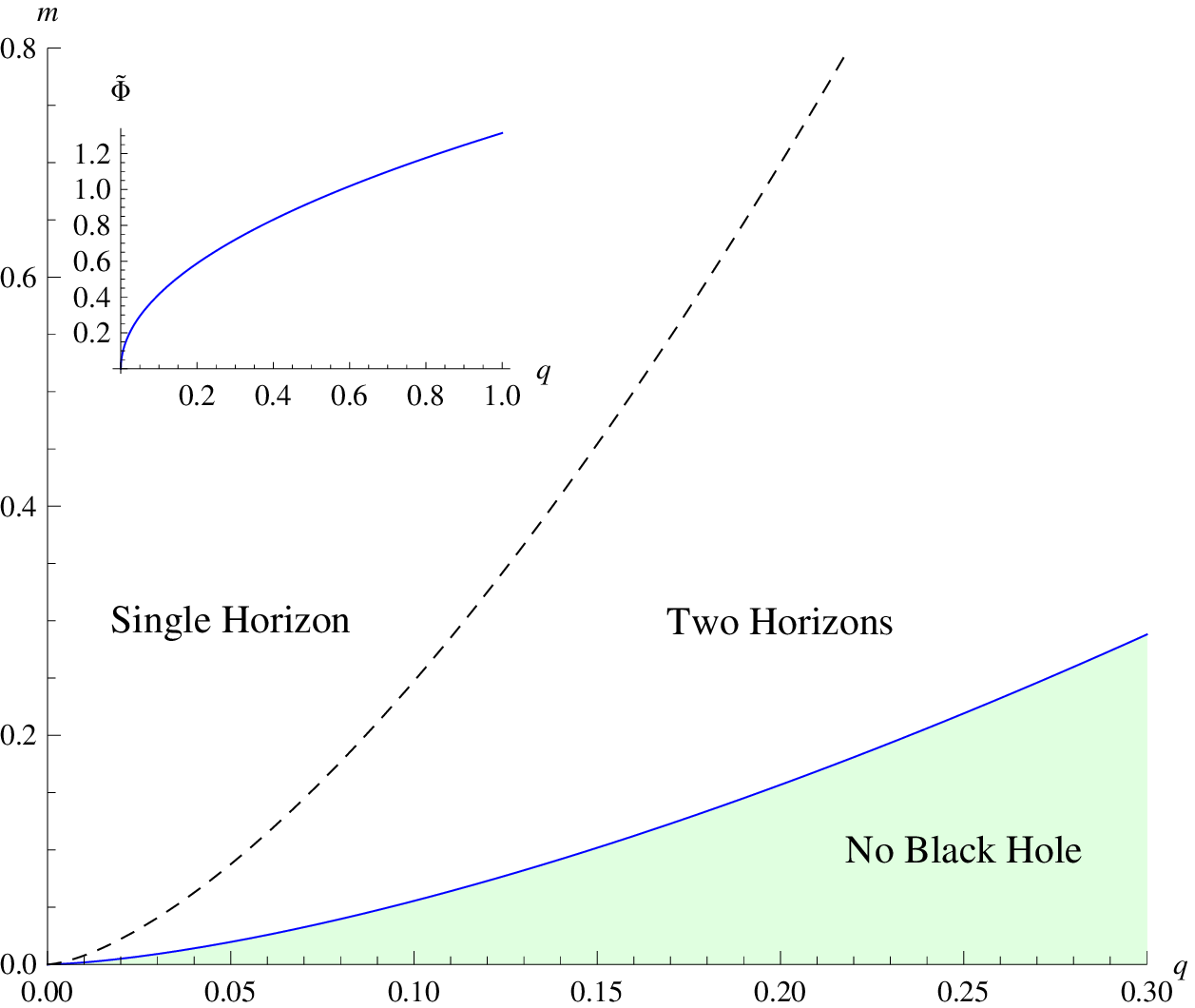}\label{fig:mq:c}}
\subfigure[{~ $k=1$ and $d=4$}]{
\includegraphics[width=0.48\textwidth]{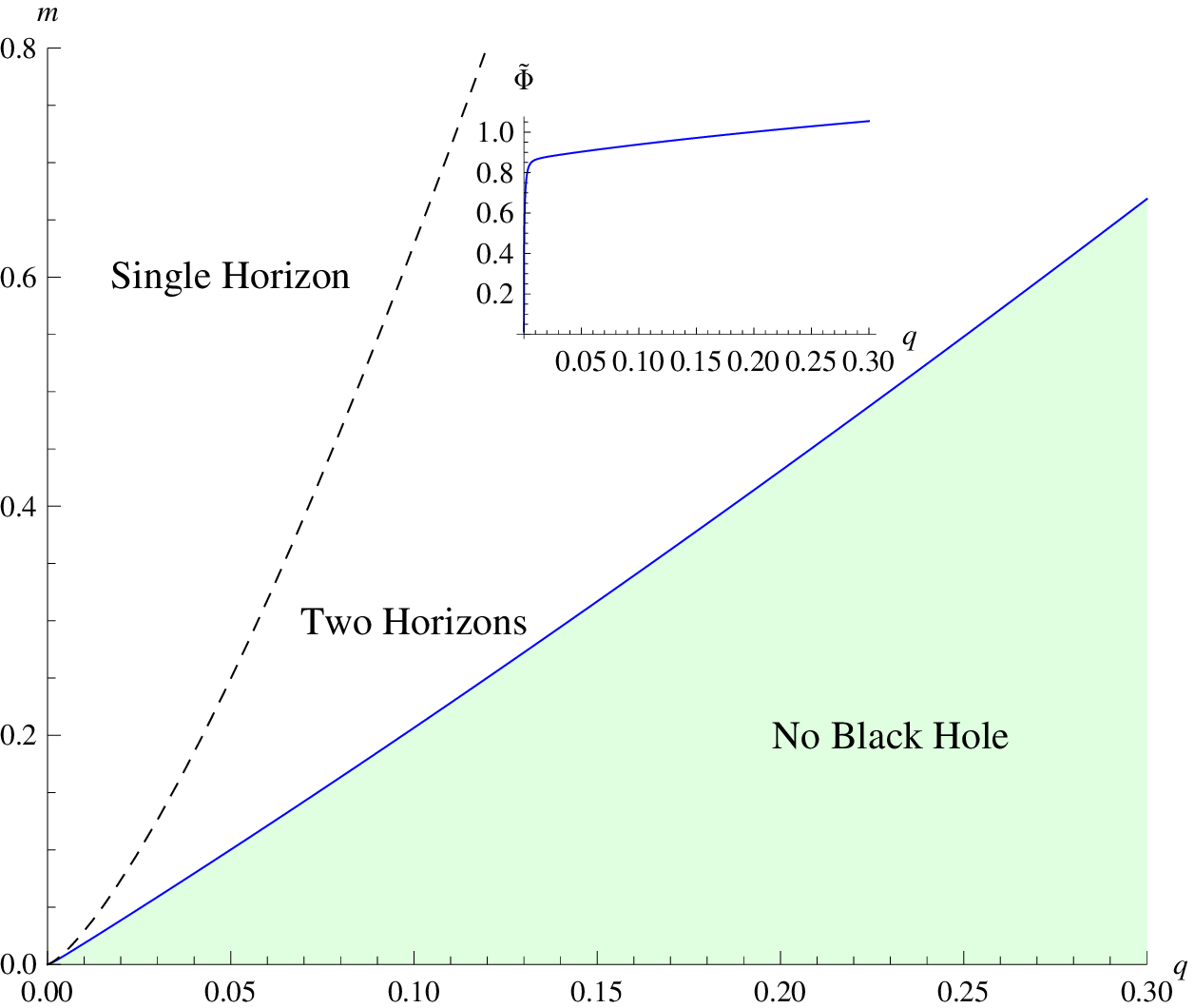}\label{fig:mq:d}}
\end{center}
\caption{The phase diagrams for Born-Infeld AdS black holes where $L=1$ and
$\beta=10$. The blue lines are extremal black holes, while the red one is some
regular spacetime with nonvanishing charges. Small figures are the plots of
the potential along the boundary lines.}%
\label{fig:mq}%
\end{figure}

The phase diagrams for Born-Infeld AdS black holes are plotted in FIG.
$\ref{fig:mq}$, for the cases with $\left\{  d=3\text{, }k=\pm1,0\right\}  $
and $\left\{  d=4\text{, }k=1\right\}  $. We also take $L=1$ and $\beta=10$ in
FIG. $\ref{fig:mq}$. The blue lines in FIG. $\ref{fig:mq}$ are extremal lines,
which are given by $m=$ $m\left(  r_{e}\left(  q\right)  ,q\right)  $. The
boundaries between the black holes with one horizon and these with two
horizons are depicted as the black dashed lines, which are given by
$m=A\left(  q\right)  $. The colored lines (red and blue) are the boundaries
between black holes and naked singularities. In FIG. $\ref{fig:mq:a}$, the red
line divides the black holes with a single horizon and the spacetime with a
naked singularity, and it meets the blue extremal line at the red dot, whose
$q\,$coordinate is $\frac{1}{2\beta}=0.05$.

To discuss the Lloyd bounds, we need to specify the electrostatic potential of
the ground states, which are the colored lines in FIG. $\ref{fig:mq}$. The
electrostatic potential at the black hole horizon, which is conjugate to the
electric charge $Q$, is \cite{BIABH-Dey:2004yt,BIABH-Cai:2004eh}
\begin{equation}
\Phi=\sqrt{\frac{d-1}{2\left(  d-2\right)  }}\frac{16\pi q}{r_{h}^{d-2}}\text{
}_{2}F_{1}\left[  \frac{d-2}{2d-2},\frac{1}{2},\frac{3d-4}{2d-2}%
,-\frac{\left(  d-1\right)  \left(  d-2\right)  q^{2}}{2\beta^{2}r_{h}^{2d-2}%
}\right]  ,\label{eq:potential}%
\end{equation}
where $r_{h}$ is the horizon's radius. When $\beta\rightarrow\infty$, the
Born-Infeld AdS black holes become the RN AdS black holes. When $k=1$ and
$d=3$, it was found \cite{IN-Brown:2015lvg} that the boundary of RN AdS black
holes in the phase diagram was the extremal line, and the potential $\Phi$
approached $16\pi$ as $\left(  q,m\right)  \rightarrow\left(  0,0\right)  $
along the extremal line. Thus for a RN AdS black hole, the ground state of the
geometry with the same electrostatic potential as this black hole is pure AdS
spacetime for $\frac{\Phi}{16\pi}\leq1$, but for $\frac{\Phi}{16\pi}>1$ it is
some extremal black hole. Now we compute the asymptotic behavior of $\Phi$ as
$\left(  q,m\right)  \rightarrow\left(  0,0\right)  $ along the boundaries:

\begin{itemize}
\item $k=0$: The boundary is the extremal line, on which $r_{e}\sim
q^{\frac{1}{d-1}}$ given by $dm\left(  r,q\right)  /dr|_{r=r_{e}}=0$. Since
$\frac{q^{2}}{r_{e}^{2d-2}}\sim1$ as $q\rightarrow0$, we find%
\begin{equation}
\Phi\sim\frac{q}{r_{e}^{d-2}}\sim q^{\frac{1}{d-1}}\rightarrow0\text{ as
}q\rightarrow0.
\end{equation}

\item $k=-1\,$: The boundary is the extremal line, on which $r_{e}\sim L$ as
$q\rightarrow0$. One then finds%
\begin{equation}
\Phi\sim\frac{q}{r_{e}^{d-2}}\sim q\rightarrow0\text{ as }q\rightarrow0.
\end{equation}
It is interesting to note that
\begin{equation}
m\rightarrow\frac{2}{d}\left(  \frac{d-2}{d}\right)  ^{\frac{d-2}{2}}%
L^{d-2}\text{ as }q\rightarrow0.
\end{equation}

\item $k=1\,$: If $d>3$, the extremal line could go to $\left(  0,0\right)  $
as $q\rightarrow0$. On the extremal line, $dm\left(  r,q\right)
/dr|_{r=r_{e}}=0$ gives that $r_{e}\sim q^{\frac{1}{d-3}}$ and $\frac{q^{2}%
}{r_{e}^{2d-2}}\rightarrow\infty$ as $q\rightarrow0$. Using eqn. $\left(
\ref{eq:potential}\right)  $, we also find%
\begin{equation}
\Phi\sim q^{\frac{1}{d-1}}\rightarrow0.
\end{equation}
If $d=3$, the boundary line around $\left(  0,0\right)  $ is the red line in
FIG. $\ref{fig:mq:a}$, on which $r_{+}=0$. Again, we have
\begin{equation}
\Phi\propto q^{\frac{1}{d-1}}\rightarrow0.
\end{equation}

\end{itemize}

Unlike the $k=1$ and $d=3$ RN AdS black holes, the potential $\Phi
\rightarrow0$ as $\left(  q,m\right)  \rightarrow\left(  0,0\right)  $ along
the boundary lines for the Born-Infeld AdS\ black holes. Thus for a
Born-Infeld AdS\ black hole with $\Phi>0$, the ground state of the geometry
with the same $\Phi$ is either some extremal black hole (blue lines) or some
regular spacetime with nonvanishing charges (red lines). In the cases with
$\left\{  d=3\text{, }k=\pm1,0\right\}  $ and $\left\{  d=4\text{,
}k=1\right\}  $, the potential along the boundary lines are plotted in FIG.
$\ref{fig:mq}$, where $\tilde{\Phi}=\frac{\Phi}{16\pi}$, $L=1$, and $\beta=10$.

\section{Holographic Conjectures of Complexity}

\label{Sec:HCC}

In this section, we will discuss CA/CV dualities for the Born-Infeld AdS black
holes. In our appendix, the action growth of the Born-Infeld AdS black holes
within the WdW patch at late-time approximation is calculated by following the
approach in \cite{IN-Lehner:2016vdi}. The action growth in the case with $k=1$
and $d=3$ was first calculated in \cite{IN-Cai:2016xho}. The growth rate of
the action $dS/dt$ depends on the number of the horizons. In fact, we find
that%
\begin{align}
\frac{dS}{dt} &  =2M-Q\Phi_{+}-\left(  d-2\right)  A\left(  q\right)
\Omega_{k,d-1}\text{ in one horizon case,}\nonumber\\
\frac{dS}{dt} &  =Q\Phi_{-}-Q\Phi_{+}\text{ in two horizons case,}%
\label{eq:CA}%
\end{align}
where $\Phi$ is the potential at the horion given by eqn. $\left(
\ref{eq:potential}\right)  $, $\Phi_{\pm}$ are $\Phi$ calculated at $r=r_{\pm
}$, and $r_{\pm}$ is the radius of the $%
\genfrac{.}{.}{0pt}{}{\text{outer}}{\text{inner}}%
$ horizon. Furthermore, CA duality indicates that, in the late time regime,%
\begin{equation}
\mathcal{\dot{C}}_{A}=\frac{1}{\pi\hbar}\frac{dS}{dt}\text{.}%
\end{equation}
On the other hand, CV duality gives \cite{IN-Couch:2016exn} that, in the late
time regime,
\begin{equation}
\mathcal{\dot{C}}_{V}=\frac{PV}{\hbar}\text{,}%
\end{equation}
where $P=d\left(  d-1\right)  /L^{2}$ is the pressure, and $V$ is the volume
of the WdW patch. For Born-Infeld AdS black holes, the rate of the complexity
at late times is then given by%
\begin{align}
\mathcal{\dot{C}}_{V} &  =\frac{\left(  d-1\right)  \Omega_{k,d-1}r_{+}^{d}%
}{L^{2}\hbar}\text{ in one horizon case,}\nonumber\\
\mathcal{\dot{C}}_{V} &  =\frac{\left(  d-1\right)  \Omega_{k,d-1}\left(
r_{+}^{d}-r_{-}^{d}\right)  }{L^{2}\hbar}\text{ in two horizons case.}%
\label{eq:CV}%
\end{align}

The Lloyd bound for a charged black hole is%
\begin{equation}
\mathcal{\dot{C}\leq}\frac{2}{\pi\hbar}\left[  \left(  M-Q\Phi\right)
-\left(  M-Q\Phi\right)  _{\text{gs}}\right]  ,\label{eq:Lbound}%
\end{equation}
where $\left(  M-Q\Phi\right)  _{\text{gs}}$ is $M-Q\Phi$ calculated in the
ground state. The ground state is on the boundary between black hole region
and no black hole region (colored lines in FIG. $\ref{fig:mq}$). If the system
is treated as a grand canonical ensemble, the ground state has the same
potential $\Phi$ as the black hole under consideration. Now we will calculate
the rate of the complexity in the CA and CV dualities and check whether the
Lloyd bound $\left(  \ref{eq:Lbound}\right)  $ is violated.

\subsection{Around Extremal Line}

We first consider a general static charged black hole with the line element%
\begin{equation}
ds^{2}=-f\left(  r\right)  dt^{2}+\frac{dr^{2}}{f\left(  r\right)  }%
+r^{2}d\Sigma_{k,d-1}^{2},
\end{equation}
where the radii of the outer and inner horizon are $r_{+}$ and $r_{-}$,
respectively. The first law of black hole thermodynamics reads%
\begin{equation}
dM=TdS+\Phi dQ\text{.}%
\end{equation}
Since the entropy $S$ is the function of $r_{+}$, one finds%
\begin{align}
\frac{\partial M\left(  r_{+},Q\right)  }{\partial Q} &  =\Phi,\nonumber\\
\frac{\partial M\left(  r_{+},Q\right)  }{\partial r_{+}} &  =T\frac
{dS}{dr_{+}}.
\end{align}
At extremality where $T=0$, we have%
\begin{equation}
\frac{\partial M\left(  r_{+},Q_{e}\right)  }{\partial r_{+}}|_{r_{+}=r_{e}%
}=0\text{,}%
\end{equation}
where $r_{e}$ and $Q_{e}$ are the radius and charge, respectively, of the
black hole at extremality. For a fixed value of $\Phi$, $r_{+}$ can be
determined by $Q\,$: $r_{+}=r_{+}\left(  Q\right)  $. Thus on the constant
$\Phi$ curve near extremality, we find%
\begin{align}
&  M\left(  r_{+}\left(  Q_{e}+\delta Q\right)  ,Q_{e}+\delta Q\right)
-\left(  Q_{e}+\delta Q\right)  \Phi-\left[  M\left(  r_{+}\left(
Q_{e}\right)  ,Q_{e}\right)  -Q_{e}\Phi\right]  \nonumber\\
&  =\left[  \left(  \frac{\partial M\left(  r_{+},Q\right)  }{\partial r_{+}%
}\frac{dr_{+}\left(  Q\right)  }{dQ}\right)  |_{Q=Q_{e}}\delta Q+\frac
{\partial M\left(  r_{+},Q\right)  }{\partial Q}|_{Q=Q_{e}}\delta Q-\Phi\delta
Q\right]  +\mathcal{O}\left(  \delta Q^{2}\right)  \sim\mathcal{O}\left(
\delta Q^{2}\right)  .
\end{align}
The Lloyd bound then becomes%
\begin{equation}
\frac{2}{\pi\hbar}\left[  \left(  M-Q\Phi\right)  -\left(  M-Q\Phi\right)
_{\text{gs}}\right]  \sim\mathcal{O}\left(  \delta Q^{2}\right)
\text{.}\label{eq:bex}%
\end{equation}
Expanding $r_{\pm}$ near extremality, we find that
\begin{equation}
r_{\pm}\approx r_{e}+c_{1}^{\pm}\delta Q,
\end{equation}
where%
\begin{align}
c_{1}^{+} &  =-\frac{\partial_{Q}\Phi\left(  r_{e},Q_{e}\right)  }%
{\partial_{r_{+}}\Phi\left(  r_{e},Q_{e}\right)  },\nonumber\\
c_{1}^{-} &  =\frac{\partial_{Q}\Phi\left(  r_{e},Q_{e}\right)  }%
{\partial_{r_{+}}\Phi\left(  r_{e},Q_{e}\right)  }-\frac{2\partial_{r}%
\Phi\left(  r_{e},Q_{e}\right)  }{\partial_{r_{+}}^{2}M\left(  r_{e}%
,Q_{e}\right)  }.
\end{align}
From these we can expand $\mathcal{\dot{C}}$ near extremality as%
\begin{align}
\mathcal{\dot{C}}_{A} &  \sim\frac{Q_{e}\partial_{r_{+}}\Phi\left(
r_{e},Q_{e}\right)  }{\pi\hbar}\left(  c_{1}^{+}-c_{1}^{-}\right)  \delta
Q,\nonumber\\
\mathcal{\dot{C}}_{V} &  \sim\frac{d\left(  d-1\right)  \Omega_{k,d-1}%
r_{e}^{d-1}}{L^{2}\hbar}\left(  c_{1}^{+}-c_{1}^{-}\right)  \delta
Q.\label{eq:cex}%
\end{align}
If $c_{1}^{+}\neq c_{1}^{-}$, the Lloyd bounds are violated near extremality
under the two proposals. For the Born-Infeld AdS\ black holes with $d=3$, we
find that%
\begin{equation}
c_{1}^{+}-c_{1}^{-}=\frac{k-\tilde{\Phi}^{2}}{\tilde{\Phi}^{2}\left(
k-2\tilde{\Phi}^{2}\right)  }-\frac{3\left(  k^{2}-8k\tilde{\Phi}^{2}%
+6\tilde{\Phi}^{4}\right)  }{10\beta^{2}L^{2}\left(  k-2\tilde{\Phi}%
^{2}\right)  ^{2}\left(  k-\tilde{\Phi}^{2}\right)  }+\mathcal{O}\left(
\beta^{-4}\right)  ,
\end{equation}
where $\tilde{\Phi}=\frac{\Phi}{16\pi}$.

\subsection{Around Regular Charged Spacetime}

As shown in FIG. $\ref{fig:mq:a}$, there is a red boundary, which is
$m=A\left(  q\right)  $ for $q\leq\frac{1}{2\beta}$, in the case with $d=3$
and $k=1$. Above this boundary, one has a black hole with a single horizon,
whose radius goes to zero as approaching the boundary. When $r\ll1$, we find%
\begin{equation}
f\left(  r\right)  =\left(  1-2q\beta\right)  -\frac{m-A\left(  q\right)  }%
{r}+\mathcal{O}\left(  r^{2}\right)  \text{,} \label{eq:fnRL}%
\end{equation}
which means that the metric is regular at $r=0$ for $m=A\left(  q\right)  $.
Therefore, one has some regular spacetime with nonvanishing charges on the red
boundary. The potential $\Phi\left(  =16\pi\tilde{\Phi}\right)  $ of the
ground states on the boundary can be obtained from finding the limit of eqn.
$\left(  \ref{eq:potential}\right)  $ as $r_{+}\rightarrow0$:%
\begin{equation}
\tilde{\Phi}=\tilde{\Phi}_{c}\sqrt{2q\beta}\leq\tilde{\Phi}_{c},
\end{equation}
where
\begin{equation}
\tilde{\Phi}_{c}=\frac{1}{\sqrt{2\pi}}\Gamma\left(  \frac{5}{4}\right)
\Gamma\left(  \frac{1}{4}\right)  \text{.}%
\end{equation}

A little bit above the boundary, the radius of a black hole with the potential
$\Phi$ is given by%
\begin{equation}
r_{+}\approx\frac{\tilde{\Phi}_{c}^{2}}{\tilde{\Phi}^{2}+\tilde{\Phi}_{c}^{2}%
}\delta m,\label{eq:rdeltam}%
\end{equation}
where $\delta m=m-m_{0}$, and $m_{0}$ is the $m$ parameter of the ground state
with the same potential $\Phi$. Since $r_{+}\ll1$ implied by eqn. $\left(
\ref{eq:rdeltam}\right)  $, eqn. $\left(  \ref{eq:fnRL}\right)  $ gives that
the temperature of the black hole is%
\begin{equation}
T\propto m-A\left(  q\right)  \text{,}%
\end{equation}
which goes to zero as approaching the ground state. For this black hole, we
find that the Lloyd bound is%
\begin{equation}
\frac{2}{\pi\hbar}\left[  \left(  M-Q\Phi\right)  -\left(  M-Q\Phi\right)
_{\text{gs}}\right]  \approx\frac{16}{\hbar}\left(  1-\frac{2\tilde{\Phi}^{2}%
}{\tilde{\Phi}^{2}+\tilde{\Phi}_{c}^{2}}\right)  \delta m.\label{eq:bRL}%
\end{equation}
On the other hand, we can expand $\mathcal{\dot{C}}$ as%
\begin{align}
\mathcal{\dot{C}}_{A} &  \approx\frac{16}{\hbar}\left(  1-\frac{3}{2}%
\frac{\tilde{\Phi}^{2}}{\tilde{\Phi}^{2}+\tilde{\Phi}_{c}^{2}}\right)  \delta
m,\nonumber\\
\mathcal{\dot{C}}_{V} &  \sim\mathcal{O}\left(  r_{+}^{3}\right)
\sim\mathcal{O}\left(  \delta m^{3}\right)  .\label{eq:cRL}%
\end{align}
It appears that the bound is satisfied in CV duality although far from
saturated near the boundary. However, the bound is violated in CA duality.

\subsection{Large $q$ on Constant $\Phi$ Curve}

Consider Born-Infeld AdS black holes with fixed potential $\Phi$. When
$q\rightarrow\infty$ along the constant $\Phi$ curve, one could have there
possibilities for $\frac{q^{2}}{r_{+}^{2d-2}}$: $\frac{q^{2}}{r_{+}^{2d-2}%
}\rightarrow0$, $\frac{q^{2}}{r_{+}^{2d-2}}\rightarrow C$ where $0<C<\infty$,
and $\frac{q^{2}}{r_{+}^{2d-2}}\rightarrow\infty$. If $\frac{q^{2}}%
{r_{+}^{2d-2}}\rightarrow\infty$, eqn. $\left(  \ref{eq:potential}\right)  $
gives that $\Phi$ $\sim q^{\frac{1}{d-1}}$ which can not be a constant.
Similarly for $\frac{q^{2}}{r_{+}^{2d-2}}\rightarrow C$, one has that
$\Phi\sim r_{+}\sim q^{\frac{1}{d-1}}$. Therefore, we could only have that
\begin{equation}
\frac{q^{2}}{r_{+}^{2d-2}}\rightarrow0\text{ as }q\rightarrow\infty\text{
along the constant }\Phi\text{ curve.}\label{eq:constant phi}%
\end{equation}
Expanding eqn. $\left(  \ref{eq:potential}\right)  $ in terms of $\frac
{q}{r_{+}^{d-1}}$ and solving it for $q$, one has that%
\begin{equation}
q\sim\sqrt{\frac{2\left(  d-2\right)  }{d-1}}\tilde{\Phi}r_{+}^{d-2}\left(
1+\frac{\left(  d-2\right)  ^{3}}{2\left(  3d-4\right)  }\frac{\tilde{\Phi
}^{2}}{\beta^{2}r_{+}^{2}}\right)  .\label{eq:constantq}%
\end{equation}
Since eqn. $\left(  \ref{eq:constant phi}\right)  $ implies that $r_{+}\gg1$
when $q\gg1$, the parameter $m$ is
\begin{equation}
m=\frac{r_{+}^{d}}{L^{2}}\left[  1+\mathcal{O}\left(  r_{+}^{-2}\right)
\right]  .\label{eq:constantm}%
\end{equation}
The Lloyd bound for $q\gg1$ $\left(  r_{+}\gg1\right)  $ is then given by%
\begin{equation}
\frac{2}{\pi\hbar}\left[  \left(  M-Q\Phi\right)  -\left(  M-Q\Phi\right)
_{\text{gs}}\right]  =\frac{2\left(  d-1\right)  \Omega_{k,d-1}}{\pi\hbar
}\frac{r_{+}^{d}}{L^{2}}\left[  1+\mathcal{O}\left(  r_{+}^{-2}\right)
\right]  .\label{eq:BLQ}%
\end{equation}

From eqns. $\left(  \ref{eq:constantq}\right)  $ and $\left(
\ref{eq:constantm}\right)  $, it follows that%
\begin{equation}
m\sim q^{\frac{d}{d-2}}\text{ for }q\gg1\text{.}\label{eq:mqL}%
\end{equation}
Since $A\left(  q\right)  \sim q^{\frac{d}{d-1}}$, the Born-Infeld AdS black
holes with fixed potential $\Phi$ always lie above the $m=A\left(  q\right)  $
line for large enough $q$, which means that these black holes always possess a
single horizon for $q\gg1$ with fixed $\Phi$. Therefore, eqns. $\left(
\ref{eq:CA}\right)  $ and $\left(  \ref{eq:CV}\right)  $ give that%
\begin{align}
\mathcal{\dot{C}}_{A} &  =\frac{2\left(  d-1\right)  \Omega_{k,d-1}}{\pi\hbar
}\frac{r_{+}^{d}}{L^{2}}\left[  1-\frac{C_{d}L^{2}\tilde{\Phi}^{\frac{d}{d-1}%
}\beta^{\frac{d-2}{d-1}}}{2\left(  d-1\right)  \Omega_{k,d-1}}r_{+}^{\frac
{-d}{d-1}}+\mathcal{O}\left(  r_{+}^{-2}\right)  \right]  ,\nonumber\\
\mathcal{\dot{C}}_{V} &  =\frac{\left(  d-1\right)  \Omega_{k,d-1}r_{+}^{d}%
}{L^{2}\hbar},\label{eq:CLQ}%
\end{align}
where%
\begin{equation}
C_{d}=\frac{2\left(  d-1\right)  }{d}\frac{\Gamma\left(  \frac{3d-4}%
{2d-2}\right)  \Gamma\left(  \frac{1}{2\left(  d-1\right)  }\right)  }%
{\sqrt{\pi}}\left(  \frac{2}{d-1}\right)  ^{\frac{d-2}{d-1}}\left[
\frac{2\left(  d-2\right)  }{d-1}\right]  ^{\frac{1}{d-1}}.
\end{equation}
We see immediately that the Lloyd bound is satisfied in CA duality for
sufficiently large $q$ and tends to be saturated as $q\rightarrow\infty$.
However in CV duality, $\mathcal{\dot{C}}$ is $\pi/2$ times as large as the
Lloyd bound for $q\gg1$.

\subsection{Numerical Results}

\begin{figure}[tb]
\begin{centering}
\includegraphics[scale=0.7]{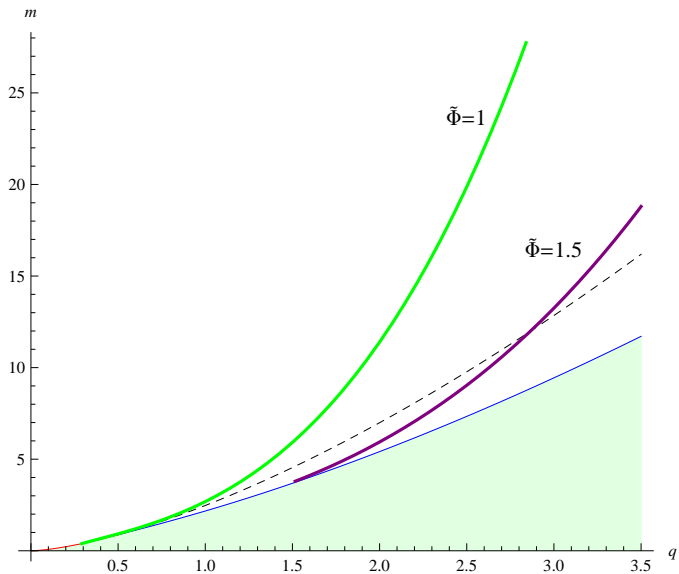}
\par\end{centering}
\caption{ Curves of constant potential $\tilde{\Phi}=1$ and $1.5\,$\ in the
case with $d=3$, $k=1$, $L=1$ and $\beta=1$. }%
\label{fig:constantphi}%
\end{figure}

\begin{figure}[tb]
\begin{center}
\subfigure[{~ Plot of $R_{A}$ versus $q$ in CA-duality along the $\tilde{\Phi}=1.5$ curve starting from
the extremal boundary.}]{
\includegraphics[width=0.48\textwidth]{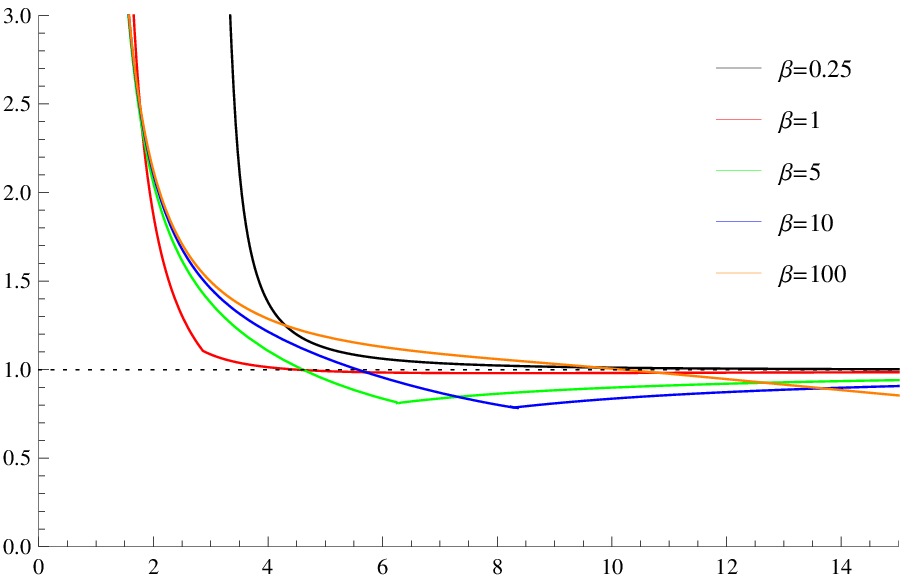}\label{fig:r:a}}
\subfigure[{~Plot of $R_{V}$ versus $q$ in CV-duality along the $\tilde{\Phi}=1.5$ curve starting from
the extremal boundary.}]{
\includegraphics[width=0.48\textwidth]{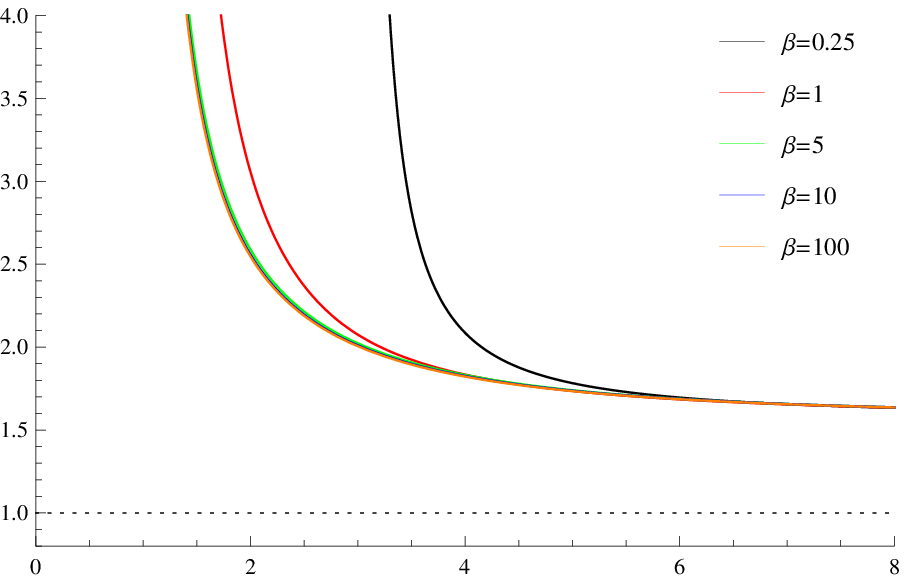}\label{fig:r:b}}
\subfigure[{~Plot of $R_{A}$ versus $q$ in CA-duality along the $\tilde{\Phi}=1$ curve starting from
the red boundary.}]{
\includegraphics[width=0.48\textwidth]{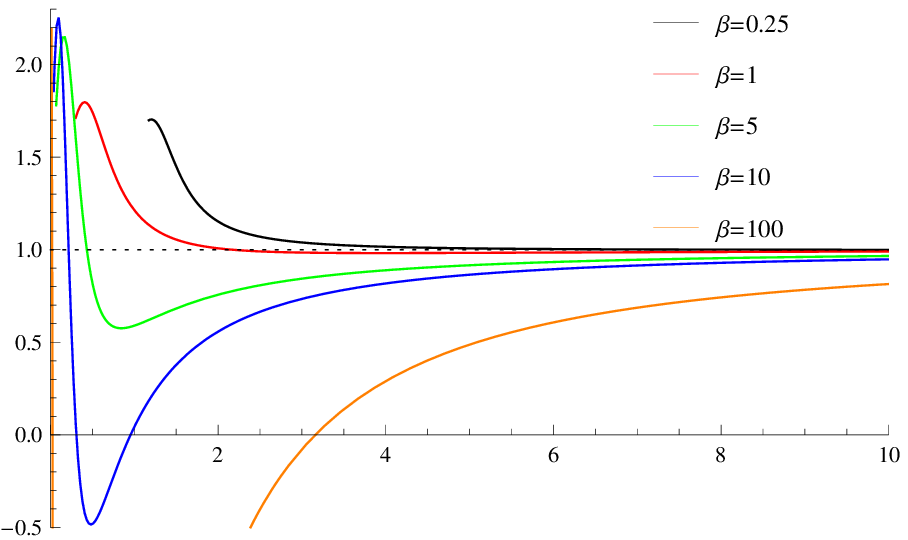}\label{fig:r:c}}
\subfigure[{~ Plot of $R_{V}$ versus $q$ in CV-duality along the $\tilde{\Phi}=1$ curve starting from
the red boundary.}]{
\includegraphics[width=0.48\textwidth]{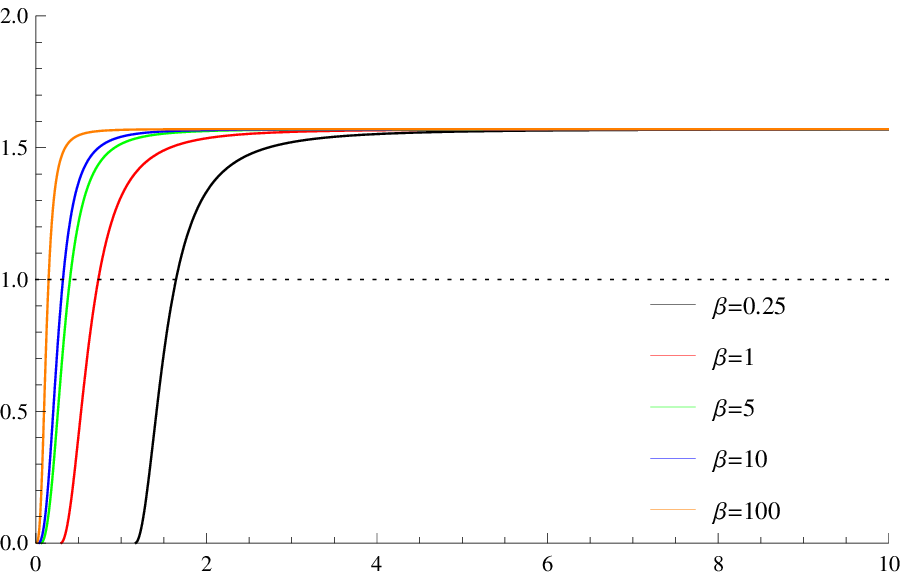}\label{fig:r:d}}
\end{center}
\caption{The rate of the complexity in CA-duality and CV-duality divided by
the Lloyd bound, $R_{A}$ and $R_{V}$ respectively, along the $\tilde{\Phi}=1$
and $\tilde{\Phi}=1.5$ curves.}%
\label{fig:r}%
\end{figure}

Here we consider two curves of constant potential, $\tilde{\Phi}=1$ and
$\tilde{\Phi}=1.5$, in the case with $d=3$ and $k=1$. These two constant
potential curves are plotted in FIG. $\ref{fig:constantphi}$ for $\beta=1$.
Note that the $\tilde{\Phi}=1$ curve (green) starts from some regular
spacetime, while the $\tilde{\Phi}=1.5$ curve (purple) starts from some
extremal black hole. Both curves enter the \textquotedblleft Single
Horizon\textquotedblright\ region for large enough $q$, which is in agreement
with the argument below eqn. $\left(  \ref{eq:mqL}\right)  $.

To check whether the Lloyd bound is violated on the curves, we define
\begin{align}
R_{A}  &  =\frac{\mathcal{\dot{C}}_{A}}{\frac{2}{\pi\hbar}\left[  \left(
M-Q\Phi\right)  -\left(  M-Q\Phi\right)  _{\text{gs}}\right]  },\nonumber\\
R_{V}  &  =\frac{\mathcal{\dot{C}}_{V}}{\frac{2}{\pi\hbar}\left[  \left(
M-Q\Phi\right)  -\left(  M-Q\Phi\right)  _{\text{gs}}\right]  }.
\end{align}
In FIG. $\ref{fig:r}$, we plot $R_{A}$ and $R_{V}$ along the $\tilde{\Phi}=1$
and $\tilde{\Phi}=1.5$ curves for $\beta=0.25$ (black), $\beta=1$ (red),
$\beta=5$ (green), $\beta=10$ (blue), and $\beta=100$ (orange). As shown in
FIG. $\ref{fig:r}$, the $R_{A}$ curves approach $R_{A}=1$ asymptotically from
below for large $q$, while the $R_{V}$ curves approach $R_{A}=\pi/2$
asymptotically, which agrees with eqns. $\left(  \ref{eq:BLQ}\right)  $ and
$\left(  \ref{eq:CLQ}\right)  $. Near extremality, $R_{A}$ and $R_{V}$ on the
$\tilde{\Phi}=1.5$ curve go to infinity as predicted by eqns. $\left(
\ref{eq:bex}\right)  $ and $\left(  \ref{eq:cex}\right)  $. When approaching
the red boundary along the $\tilde{\Phi}=1$ curve, $R_{A}$ and $R_{V}$ go
above $R_{A}=1$ and to zero, respectively, which also agrees with eqns.
$\left(  \ref{eq:bRL}\right)  $ and $\left(  \ref{eq:cRL}\right)  $.

Along the $\tilde{\Phi}=1.5$ curve, FIG. $\ref{fig:r:a}$ shows that the Lloyd
bound is satisfied in CA duality for large enough $q$, while FIG.
$\ref{fig:r:b}$ shows that the Lloyd bound is violated in CV duality. Note
that the kinks in the $R_{A}$ curves in FIG. $\ref{fig:r:a}$ are where the
$\tilde{\Phi}=1.5$ curve enter the \textquotedblleft Single
Horizon\textquotedblright\ region from the "Two Horizons" region. Along the
$\tilde{\Phi}=1$ curve, FIG. $\ref{fig:r:d}$ shows that the Lloyd bound is
only satisfied in CV duality for small $q$. It is interesting to see that the
$R_{A}$ curves in FIG. $\ref{fig:r:c}$ start to oscillate for small $q$ when
$\beta$ is large enough ($\beta=5$, $10$, and $100$). Even for $\beta=10$ and
$100$, there is a range of $q$ over which $\mathcal{\dot{C}}_{A}<0$.

In summary, the Lloyd bound is violated in CA duality as we approach the
ground states, but this bound tend to be saturated as we go away from the
ground states with fixed potential. As noted in \cite{IN-Brown:2015lvg}, the
violations near the ground states have something to do with hair. In CV
duality, the Lloyd bound is violated everywhere along the constant potential
curves, except near the ground states on the red line.

\section{Discussion and Conclusion}

\label{Sec:Con}

In this paper, we first obtained the phase diagram of Born-Infeld AdS black
holes and then checked whether the Lloyd bound was violated in CA and CV
dualities. In section \ref{Sec:BIABH}, we showed that the Born-Infeld black
hole solution could possess a naked singularity, a single horizon, or two
horizons, depending on the values of its parameters $q$ and $m$. Except the
$k=$ $1$ and $d=3$ case, the boundaries between \textquotedblleft Black
Hole\textquotedblright\ region and \textquotedblleft No Black
Hole\textquotedblright\ region were extremal lines (blue lines in FIG.
$\ref{fig:mq}$). However in the $k=$ $1$ and $d=3$ case, there was an
additional boundary (red line in FIG. $\ref{fig:mq:a}$), on which were some
regular spacetime with nonvanishing charges. It is noteworthy that unlike a RN
AdS black hole, the ground state of a Born-Infeld AdS black hole with
potential $\Phi>0$ could not be the empty AdS spacetime.

In section \ref{Sec:HCC}, we calculated the Lloyd bound and the rate of the
complexity at late times in CA and CV dualities near the boundaries and for
large $q$ on the constant $\Phi$ curves. The results of whether the Lloyd
bound was violated are summarized in TABLE \ref{tab:result}. We also found
that for a general static charged AdS black hole with the charge $Q$ near
extremality, the Lloyd bound in eqn. $\left(  \ref{eq:Lbound}\right)  $ was
always $\mathcal{O}\left(  \delta Q^{2}\right)  $, where $\delta Q\equiv
Q-Q_{e}$, and $Q_{e}$ was the charge of the extremal black hole with the same
potential. If the difference between the outer and inner horizon radii is
$\mathcal{O}\left(  \delta Q\right)  $, which is the case for RN AdS and
Born-Infeld AdS black holes, then the Lloyd bound is usually violated near
extremality. \begin{table}[h]
\begin{center}
$%
\begin{tabular}
[c]{|c|c|c|c|}\hline
& Near Extremal Line & Near Red Line & Large $q$ on Constant $\Phi$
Curve\\\hline
CA duality & Violated & Violated & Tend to be Saturated\\\hline
CV duality & Violated & Satisfied & Violated\\\hline
\end{tabular}
\ $
\end{center}
\caption{Check of whether Lloyd bound is violated or satisfied.}%
\label{tab:result}%
\end{table}

In the $d=3$ and $k=1$ case, we plotted the rate of the complexity in CA and
CV dualities divided by the Lloyd bound along the $\tilde{\Phi}=1$ and
$\tilde{\Phi}=1.5$ curves in FIG. $\ref{fig:r}$. It appears that the Lloyd
bound in CA duality was violated near the ground states but tended to be
saturated as moving away from the ground states along the constant $\Phi$
curves. On the other hand, the Lloyd bound in CV duality was violated along
the constant $\Phi$ curves, except near the ground states on the red line.
Since the hair may play a role in the violations near the ground states, it
seems from these observations that CA duality is slightly favored.

Finally, we want to briefly discuss the differences between our results and
these of RN AdS black holes. The ground state of a RN AdS black hole is either
the empty AdS space or\ some extremal black hole. However, the ground state of
a Born-Infeld AdS black hole is either some charged regular spacetime
or\ extremal black hole, but could not be the empty AdS space as long as the
potential is not zero. As shown by eqn. $\left(  5.15\right)  $ in
\cite{IN-Couch:2016exn} and FIG. $6$ in \cite{IN-Brown:2015lvg}, if the ground
state was the empty AdS space, $\mathcal{\dot{C}}_{A}$ for a RN AdS black hole
with $d=3$ and $k=1$ always violated the Lloyd bound $\left(  \ref{eq:LBoundQ}%
\right)  $ along a constant potential curve, even when $q\rightarrow\infty$.
However for a Born-Infeld AdS black hole, our results show that the Lloyd
bound in CA duality is satisfied for large enough $q$ along a constant
potential curve. When $q$ is very large with fixed potential, we have obtained
$\frac{q}{r_{+}^{d-1}}\ll1$, and the metric in eqns. $\left(  \ref{eq:BIBH}%
\right)  $ is almost the same as that of a RN AdS black hole outside the outer
horizon. In this case, physics over the region outside the outer horizon of
the Born-Infeld AdS black hole does not differ much from that of the RN AdS
black hole. Different behavior of $\mathcal{\dot{C}}_{A}$ for RN AdS and
Born-Infeld AdS black holes with large $q$ on the constant potential curves
means that the complexity encodes physics behind black hole horizons.

\begin{acknowledgments}
We are grateful to Song He, Houwen Wu, and Zheng Sun for useful discussions.
This work is supported in part by NSFC (Grant No. 11005016, 11175039 and 11375121).
\end{acknowledgments}

\appendix

\section{Rate of Action of Born-Infeld AdS Black Holes}

\begin{figure}[tb]
\begin{center}
\subfigure[{~ Single Horizon Case}]{
\includegraphics[width=0.48\textwidth]{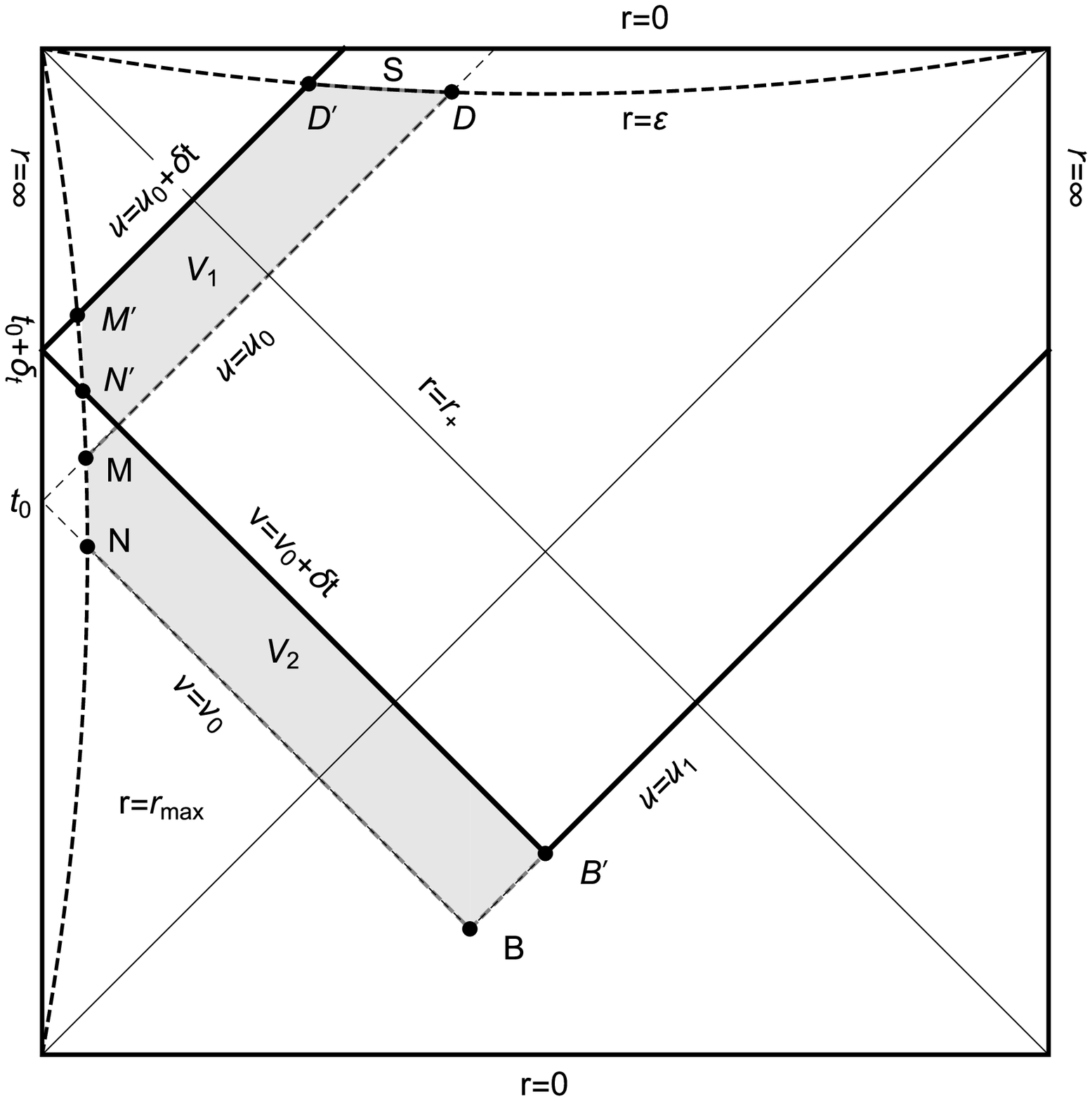}\label{fig:prdiagram:a}}
\subfigure[{~Two Horizons Case}]{
\includegraphics[width=0.48\textwidth]{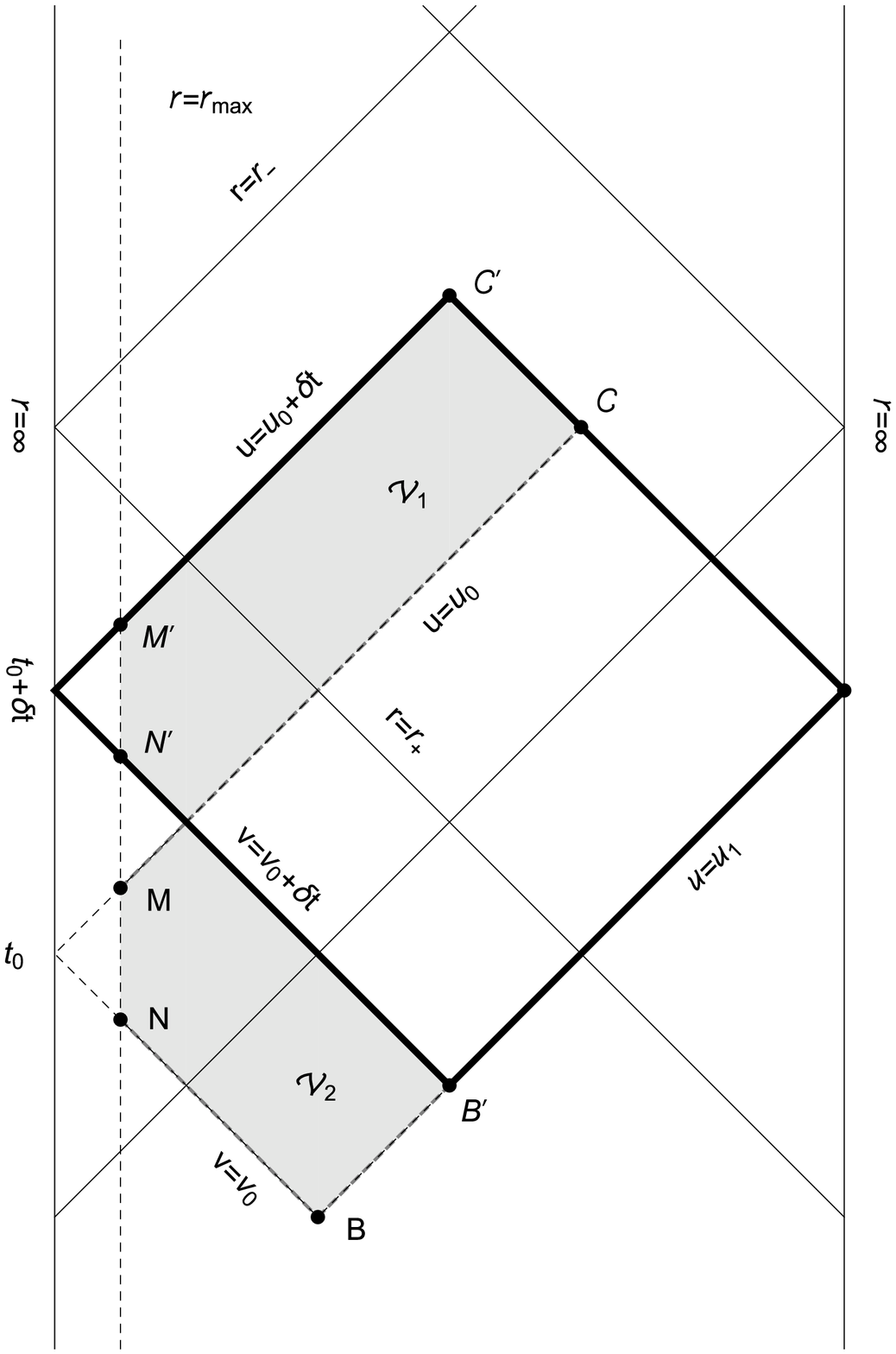}\label{fig:prdiagram:b}}
\end{center}
\caption{Wheeler-deWitt patches of Born-Infeld AdS black holes at $t_{L}%
=t_{0}$ and $t_{L}=t_{0}+\delta t$. The lines $r=r_{\max}$ and $r=\varepsilon$
are cut-off surfaces.}%
\label{fig:prdiagram}%
\end{figure}

In this appendix, we use the methods in \cite{IN-Lehner:2016vdi} to calculate
the change of action, $\delta S=S\left(  t_{0}+\delta t\right)  -S\left(
t_{0}\right)  $, of the Wheeler-DeWitt patch at late times. The Penrose
diagrams for two-sided eternal Born-Infeld AdS black holes are illustrated in
FIG. $\ref{fig:prdiagram}\,$, along with the Wheeler-DeWitt patches at
$t=t_{0}$ and $t_{0}+\delta t$. Here we fix the time on the right boundary and
only vary it on the left boundary. There is a divergence appearing when
calculating the action near the boundary $r=\infty$. So a surface of constant
$r=r_{\max}$ is defined to regulate the action. In \cite{IN-Cai:2017sjv}, the
action was regulated by defining the boundaries of the WdW patch originate
slightly inside the AdS boundary. It turns out that these two choices for the
regulator yield the same results. We also introduce a spacelike surface
$r=\varepsilon$ near the future singularities and let $\varepsilon
\rightarrow0$ at the end of calculations. Note that we have an affine
parametrization for each null surface, and these make no contribution to the
action. To calculate $\delta S$, we introduce the null coordinates $u$ and $v$
in the metric $\left(  \ref{eq:BIBH}\right)  $:%
\begin{align}
u &  =t-r^{\ast}\nonumber\\
v &  =t+r^{\ast},
\end{align}
where
\begin{equation}
r^{\ast}=\int f^{-1}\left(  r\right)  dr.
\end{equation}

\subsection{Single Horizon Case}

We calculate $\delta S$ for a Born-Infeld AdS black hole with a single
horizon, whose Penrose diagram is illustrated in FIG. $\ref{fig:prdiagram:a}$.
Due to time translation, the joint contributions from $\mathcal{D}$ and
$\mathcal{D}^{\prime}$ are identical, and they therefore make no contribution
to $\delta S$. Similarly, the joint and surface contributions from
$\mathcal{MN}$ cancel against these from $\mathcal{M}^{\prime}\mathcal{N}%
^{\prime}$ on $r=r_{\max}$ in calculating $\delta S$. Therefore, we have%
\begin{equation}
\delta S=S_{\mathcal{V}_{1}}-S_{\mathcal{V}_{2}}+2\int_{\mathcal{S}}%
d^{d}x\sqrt{\left\vert h\right\vert }K+2\int_{\mathcal{B}^{\prime}}%
d^{d-1}x\sqrt{\sigma}a-2\int_{\mathcal{B}}d^{d-1}x\sqrt{\sigma}a,
\label{eq:deltaS}%
\end{equation}
where we follow the conventions in \cite{IN-Carmi:2016wjl}.

Using the Born-Infeld AdS black hole solution $\left(  \ref{eq:BIBH}\right)
$, we find that the volume contribution is%
\begin{equation}
S_{\mathcal{V}}=\Omega_{k,d-1}\int_{\mathcal{V}}d\omega F\left(  r\right)
,\label{eq:SV}%
\end{equation}
where $\omega=\left\{  u,v\right\}  $, and%
\begin{equation}
F\left(  r\right)  =2r^{d-2}\left(  k-\frac{m}{r^{d-2}}-f\left(  r\right)
\right)  \text{.}%
\end{equation}
The region $\mathcal{V}_{1}$ is bounded by the null surfaces $u=u_{0}$,
$u=u_{0}+$ $\delta t$, $v=v_{0}+$ $\delta t$, the spacelike surface
$r=\varepsilon$, and the timelike surface $r=r_{\max}$. Using eqn. $\left(
\ref{eq:SV}\right)  $, we have that%
\begin{equation}
S_{\mathcal{V}_{1}}=\Omega_{k,d-1}\int_{u_{0}}^{u_{0}+\delta t}duF\left(
r\right)  |_{\varepsilon}^{\min\left\{  r_{\max},\rho\left(  u\right)
\right\}  }dr,
\end{equation}
where $r^{\ast}\left(  \rho\left(  u\right)  \right)  =\left(  v_{0}+\delta
t-u\right)  /2$. Since $\lim_{r\rightarrow0}F\left(  r\right)  =-2A\left(
q\right)  $, we find that%
\begin{equation}
S_{\mathcal{V}_{1}}=\Omega_{k,d-1}\int_{u_{0}}^{u_{0}+\delta t}du\left[
F\left(  r\right)  |_{r=\min\left\{  r_{\max},\rho\left(  u\right)  \right\}
}+2A\left(  q\right)  \right]  ,
\end{equation}
where $A\left(  q\right)  $ is given by eqn. $\left(  \ref{eq:Aq}\right)  $.
Similarly for $\mathcal{V}_{2}$, one has that%
\begin{equation}
S_{\mathcal{V}2}=\Omega_{k,d-1}\int_{v_{0}}^{v_{0}+\delta t}dvF\left(
r\right)  |_{\rho_{1}\left(  v\right)  }^{\min\left\{  r_{\max},\rho
_{0}\left(  v\right)  \right\}  },\label{eq:SV2S}%
\end{equation}
where $r^{\ast}\left(  \rho_{0/1}\left(  v\right)  \right)  =\left(
v-u_{0/1}\right)  /2$. Performing the change of variables $u=u_{0}%
+v_{0}+\delta t-v$, we have that%
\begin{equation}
\int_{v_{0}}^{v_{0}+\delta t}dvF\left(  r\right)  |_{r=\min\left\{  r_{\max
},\rho_{0}\left(  v\right)  \right\}  }=\int_{u_{0}}^{u_{0}+\delta
t}duF\left(  r\right)  |_{r=\min\left\{  r_{\max},\rho\left(  u\right)
\right\}  },
\end{equation}
and hence%
\begin{equation}
S_{\mathcal{V}_{1}}-S_{\mathcal{V}_{2}}=\Omega_{k,d-1}\left[  \int_{v_{0}%
}^{v_{0}+\delta t}dvF\left(  r\right)  |_{r=\rho_{1}\left(  v\right)
}+2A\left(  q\right)  \int_{u_{0}}^{u_{0}+\delta t}du\right]  .\label{eq:S1}%
\end{equation}
At late times, one has that $\rho_{1}\left(  v\right)  \approx r_{+}$, and%
\begin{equation}
S_{\mathcal{V}_{1}}-S_{\mathcal{V}_{2}}=\Omega_{k,d-1}\left[  F\left(
r_{h}\right)  +2A\left(  q\right)  \right]  \delta t\text{.}%
\end{equation}

There is a timelike hypersurface at $r=\varepsilon$, with outward-directed
normal vectors from the region of interest. The normal vector is%
\begin{equation}
n_{\mu}dx^{\mu}=\frac{-1}{\sqrt{-f\left(  r\right)  }}dr\text{.}%
\end{equation}
The trace of extrinsic curvature is
\begin{equation}
K=\frac{1}{r^{d-1}}\partial_{r}\left(  r^{d-1}\sqrt{-f\left(  r\right)
}\right)  .
\end{equation}
Therefore, the surface contributions from $r=\varepsilon$ is%
\begin{equation}
2\int_{\mathcal{S}}d^{d}x\sqrt{\left\vert h\right\vert }K=2\left(  m-A\left(
q\right)  \right)  \Omega_{k,d-1}\frac{\delta t}{r^{d/2-1}}\partial_{r}\left(
r^{d/2}\right)  |_{r=\varepsilon}=\left[  m-A\left(  q\right)  \right]
d\Omega_{k,d-1}\delta t, \label{eq:S2}%
\end{equation}
where we use $\sqrt{\left\vert h\right\vert }=\sqrt{-f\left(  r\right)
}r^{d-1}d\Omega_{k,d-1}$.

Following \cite{IN-Carmi:2016wjl}, the integrand $a$ in the joint terms of
eqn. $\left(  \ref{eq:deltaS}\right)  $ is
\begin{align}
a  &  =\epsilon\ln\left\vert \mathbf{k}_{1}\cdot\mathbf{k}_{2}/2\right\vert
,\nonumber\\
\epsilon &  =-\text{sign}\left(  \mathbf{k}_{1}\cdot\mathbf{k}_{2}\right)
\text{sign}\left(  \mathbf{\hat{k}}\cdot\mathbf{k}_{2}\right)  ,
\end{align}
where for $\mathcal{B}$ and $\mathcal{B}^{\prime}$,
\begin{align}
\left(  \mathbf{k}_{1}\right)  _{\mu}  &  =-c_{1}\partial_{\mu}\left(
t+r^{\ast}\right)  ,\nonumber\\
\left(  \mathbf{k}_{2}\right)  _{\mu}  &  =c_{2}\partial_{\mu}\left(
t-r^{\ast}\right)  ,
\end{align}
and the auxiliary null vectors $\mathbf{\hat{k}}$ is the null vector
orthogonal to the joint and pointing outward from the boundary region.
Therefore, we find that%
\begin{equation}
2\int_{\mathcal{B}^{\prime}}d^{d-1}x\sqrt{\sigma}a-2\int_{\mathcal{B}}%
d^{d-1}x\sqrt{\sigma}a=2\Omega_{k,d-1}\left[  h\left(  r_{\mathcal{B}^{\prime
}}\right)  -h\left(  r_{\mathcal{B}}\right)  \right]  ,
\end{equation}
where%
\begin{equation}
h\left(  r\right)  =r^{d-1}\ln\left(  -\frac{f\left(  r\right)  }{c_{1}c_{2}%
}\right)  .
\end{equation}
At late times, we have that $r_{\mathcal{B}}\approx r_{+}$ and
\begin{equation}
h\left(  r_{B^{\prime}}\right)  -h\left(  r_{B}\right)  =\frac{f\left(
r\right)  }{2}\frac{dh\left(  r\right)  }{dr}|_{r=r_{B}}\delta t=\frac{1}%
{2}r^{d-1}\frac{df\left(  r\right)  }{dr}|_{r=r_{+}}\delta t,
\end{equation}
where we use $dr=f\left(  r\right)  \delta t/2$ on $u=u_{1}$. Thus, this gives%
\begin{equation}
2\int_{\mathcal{B}^{\prime}}d^{d-1}x\sqrt{\sigma}a-2\int_{\mathcal{B}}%
d^{d-1}x\sqrt{\sigma}a=\Omega_{k,d-1}r_{+}^{d-1}f^{\prime}\left(
r_{+}\right)  \delta t. \label{eq:S3}%
\end{equation}

Combining eqns. $\left(  \ref{eq:S1}\right)  $, $\left(  \ref{eq:S2}\right)
$, and $\left(  \ref{eq:S3}\right)  $, we arrive at%
\begin{equation}
\frac{dS}{dt}=2M-Q\Phi_{+}-\left(  d-2\right)  A\left(  q\right)
\Omega_{k,d-1} \label{eq:ds/dtS}%
\end{equation}
where we use $f\left(  r_{+}\right)  =0$, and $\Phi_{+}$ is the potential
$\Phi$ evaluated at $r=r_{+}$. When $k=1$ and $d=3$, eqn. $\left(
\ref{eq:ds/dtS}\right)  $ becomes%
\begin{equation}
\frac{dS}{dt}=2M-Q\Phi_{+}-16\pi\beta^{1/2}Q^{3/2}\frac{\Gamma\left(
1/4\right)  \Gamma\left(  5/4\right)  }{3\Gamma\left(  1/2\right)  },
\label{eq:ds/dt31}%
\end{equation}
where $Q=q$ in the $k=1$ and $d=3$ case. Taking into account that $16\pi G=1$
in our paper and $G=1$ in \cite{IN-Cai:2017sjv}, our result $\left(
\ref{eq:ds/dt31}\right)  $ agrees with eqn. (3.26) in \cite{IN-Cai:2017sjv}.

\subsection{Two Horizons Case}

The Penrose diagram for a Born-Infeld AdS black hole with two horizons is
illustrated in FIG. $\ref{fig:prdiagram:b}$. Thus, we have
\begin{equation}
\delta S=S_{\mathcal{V}_{1}}-S_{\mathcal{V}_{2}}+2\int_{\mathcal{B}^{\prime}%
}d^{d-1}x\sqrt{\sigma}a-2\int_{\mathcal{B}}d^{d-1}x\sqrt{\sigma}%
a+2\int_{\mathcal{C}^{\prime}}d^{d-1}x\sqrt{\sigma}a-2\int_{\mathcal{C}%
}d^{d-1}x\sqrt{\sigma}a.
\end{equation}
While the volume contribution $S_{\mathcal{V}_{2}}\,$is also given by eqn.
$\left(  \ref{eq:SV2S}\right)  $, we find that, in this case,
\begin{equation}
S_{\mathcal{V}_{1}}=\Omega_{k,d-1}\int_{u_{0}}^{u_{0}+\delta t}duF\left(
r\right)  |_{\tilde{\rho}_{1}\left(  u\right)  }^{\min\left\{  r_{\max}%
,\rho\left(  u\right)  \right\}  }dr,
\end{equation}
where
\begin{equation}
r^{\ast}\left(  \tilde{\rho}_{1}\left(  u\right)  \right)  =\frac{v_{1}-u}{2}.
\end{equation}
Hence the volume contribution to $\delta S$ is
\begin{align}
S_{\mathcal{V}_{1}}-S_{\mathcal{V}_{2}}  &  =\Omega_{k,d-1}\left[  \int
_{v_{0}}^{v_{0}+\delta t}dvF\left(  r\right)  |_{r=\rho_{1}\left(  v\right)
}-\int_{u_{0}}^{u_{0}+\delta t}duF\left(  r\right)  |_{r=\tilde{\rho}%
_{1}\left(  u\right)  }dr\right] \nonumber\\
&  =\Omega_{k,d-1}\left[  F\left(  r_{+}\right)  -F\left(  r_{-}\right)
\right]  \delta t,
\end{align}
where the portion of $\mathcal{V}_{1}$ below the future horizon cancels
against the portion of $\mathcal{V}_{2}$ above the past horizon. The joint
contributions from $\mathcal{B}$ and $\mathcal{B}^{\prime}$ are the same as in
the case with a single horizon. Analogously to calculating the joint
contributions from $\mathcal{B}$ and $\mathcal{B}^{\prime}$, we find that%
\begin{equation}
2\int_{\mathcal{C}^{\prime}}d^{d-1}x\sqrt{\sigma}a-2\int_{\mathcal{C}}%
d^{d-1}x\sqrt{\sigma}a=-\Omega_{k,d-1}r_{-}^{d-1}f^{\prime}\left(
r_{-}\right)  \delta t,
\end{equation}
where $r_{-}$ is the inner horizon radius. Summing up all the contributions,
we obtains that%
\begin{equation}
\frac{dS}{dt}=Q\Phi_{-}-Q\Phi_{+}, \label{eq:ds/dtD}%
\end{equation}
where $\Phi_{\pm}$ is the potential $\Phi$ evaluated at $r=r_{\pm}$. When
approaching the boundary between the \textquotedblleft Single
Horizon\textquotedblright\ and \textquotedblleft Two
Horizons\textquotedblright\ regions, we have $r_{-}\rightarrow0$ and
$Q\Phi_{-}\rightarrow A\left(  q\right)  d\Omega_{k,d-1}$. Since $m=A\left(
q\right)  $ on this boundary, eqn. $\left(  \ref{eq:ds/dtD}\right)  $ becomes
\begin{equation}
\frac{dS}{dt}\rightarrow2M-Q\Phi_{+}-\left(  d-2\right)  A\left(  q\right)
\Omega_{k,d-1}.
\end{equation}
Comparing with eqn. $\left(  \ref{eq:ds/dtS}\right)  $, we find that $dS/dt$
is continuos when crossing this boundary.

\end{document}